\documentclass[twocolumn,floatfix,aps,prb,eqsecnum,superscriptaddress]{revtex4}
\usepackage{graphicx}
\usepackage{amsmath}
\usepackage{amssymb}
\usepackage{bm}

\usepackage{color}

\begin{document}

\title{Two-dimensional Josephson vortex lattice and anomalously slow decay of the Fraunhofer oscillations in a ballistic SNS junction with a warped Fermi surface}
\author{V. P. Ostroukh}
\affiliation{Instituut-Lorentz, Universiteit Leiden, P.O. Box 9506, 2300 RA Leiden, The Netherlands}
\author{B. Baxevanis}
\affiliation{Instituut-Lorentz, Universiteit Leiden, P.O. Box 9506, 2300 RA Leiden, The Netherlands}
\author{A. R. Akhmerov}
\affiliation{Kavli Institute of Nanoscience, Delft University of Technology, P.O. Box 4056, 2600 GA Delft, The Netherlands}
\author{C. W. J. Beenakker}
\affiliation{Instituut-Lorentz, Universiteit Leiden, P.O. Box 9506, 2300 RA Leiden, The Netherlands}
\date{June 2016}
\begin{abstract}
The critical current of a Josephson junction is an oscillatory function of the enclosed magnetic flux $\Phi$, because of quantum interference modulated with periodicity $h/2e$. We calculate these Fraunhofer oscillations in a two-dimensional (2D) ballistic superconductor--normal-metal--superconductor (SNS) junction. For a Fermi circle the amplitude of the oscillations decays as $1/\Phi$ or faster. If the Fermi circle is strongly warped, as it is on a square lattice near the band center, we find that the amplitude decays slower $\propto 1/\sqrt\Phi$ when the magnetic length $l_m=\sqrt{\hbar/eB}$ drops below the separation $L$ of the NS interfaces. The crossover to the slow decay of the critical current is accompanied by the appearance of a 2D array of current vortices and antivortices in the normal region, which form a bipartite rectangular lattice with lattice constant $\simeq l_m^2/L$. The 2D lattice vanishes for a circular Fermi surface, when only the usual single row of Josephson vortices remains.
\end{abstract}
\maketitle

\section{Introduction}
\label{intro}

A junction between two superconductors responds to an imposed magnetic flux $\Phi$ by producing a chain of circulating current vortices, known as Josephson vortices.\cite{Tinkham} The critical current $I_{c}(\Phi)$ oscillates with period $\Phi_0=h/2e$ and amplitude $\propto\Phi_0/\Phi$. These socalled Fraunhofer oscillations are a macroscopic quantum interference effect, first observed in 1963 in a tunnel junction.\cite{Row63} The effect is now used as a sensitive probe of ballistic transport and edge currents in graphene and topological insulators.\cite{Har14,Pri15,Cal15,Har15,All16,Ben16}

Since the self-field of the current vortices is typically too weak to screen the imposed magnetic field $B$ from the junction area, the arrangement of Josephson vortices is governed by quantum interference --- unaffected by the classical electrostatics that governs the two-dimensional (2D) Abrikosov vortex lattice in the bulk superconductor.\cite{Tinkham} The fundamental question addressed here, is whether quantum interference by itself is capable of producing a 2D vortex lattice in a Josephson junction. It is known that the linear arrangement of the vortices along the superconducting interface is modified by insulating boundaries,\cite{Hei98,Led99,Bar99,Kim15} in a junction of lateral width $W$ comparable to the separation $L$ of the interfaces. But in wide junctions ($W\gg L$), when boundary effects are irrelevant, only linear arrangements of Josephson vortices are known.\cite{Cue07,Ber08,Ali12,Ali15,Amu16}

We have discovered that a 2D Josephson vortex lattice appears when the circular Fermi surface acquires a square or hexagonal distortion. Such a warped Fermi surface has flattened facets that produce a nonisotropic velocity distribution of the conduction electrons, peaked at  velocity directions normal to the facets. Analytical and numerical calculations of the supercurrent distribution in the high-field regime (magnetic length $l_m=\sqrt{\hbar/eB}$ less than $L$) reveal the appearance of multiple rows of vortex-antivortex pairs, forming a 2D bipartite rectangular lattice in the normal region with lattice constant 
\begin{equation}
a_{\rm vortex}=\frac{W\Phi_0}{\Phi}=\frac{\pi l_m^2}{L}.\label{avortexdef}
\end{equation}

\begin{figure}[tb]
\centerline{\includegraphics[width=1\linewidth]{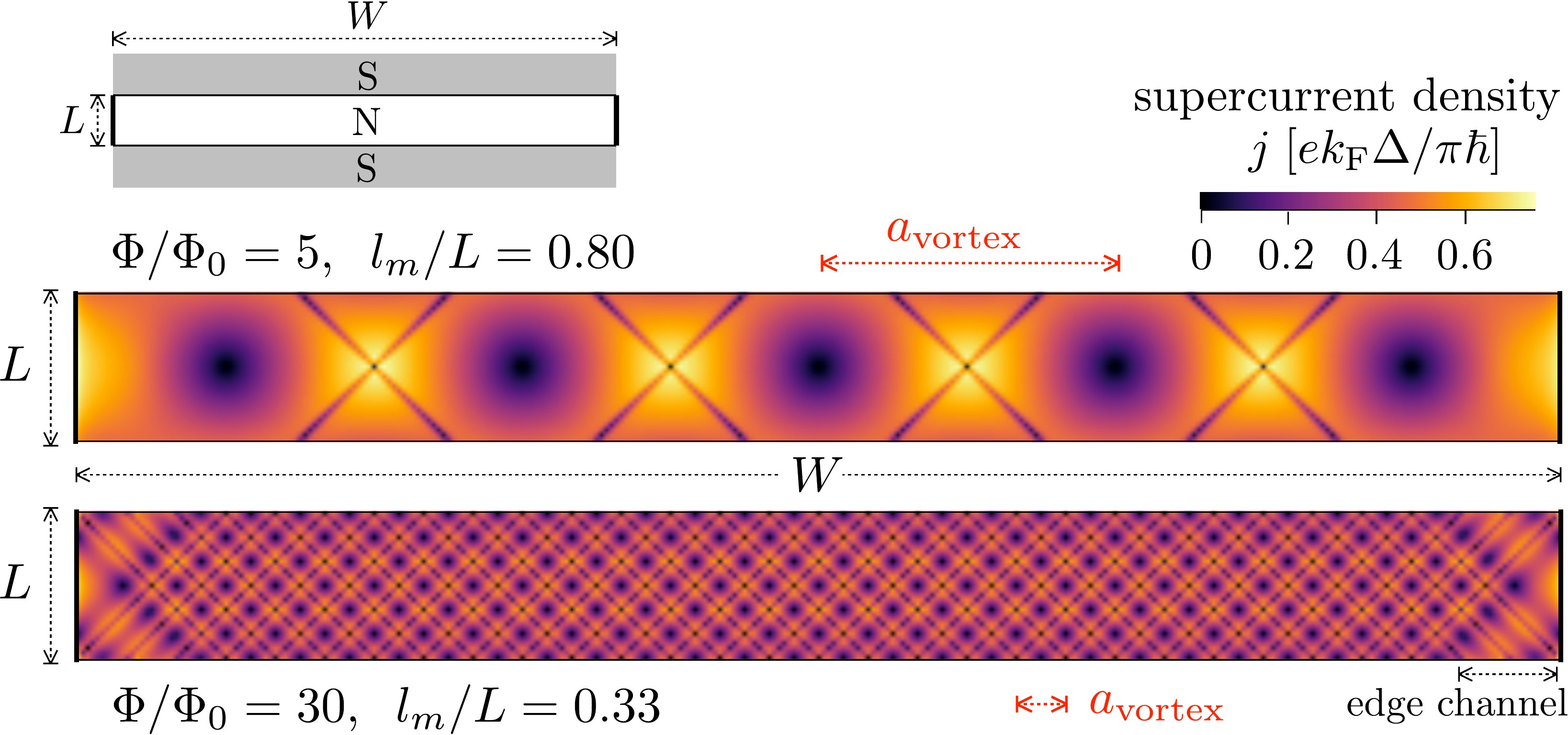}}
\caption{Supercurrent density in an SNS (superconductor--normal-metal--superconductor) Josephson junction, resulting from the numerical simulation of Sec.\ \ref{numerics} on a square lattice with a half-filled band and a square Fermi surface (lattice constant $a_0$, normal region of size $W=10L=300\,a_0$, band width $2E_0$, Fermi velocity $v_{\rm F}\equiv E_0a_0/\sqrt{2}\hbar$, resulting in $N=282$ transverse modes per spin direction at the Fermi level, superconducting gap $\Delta=2.5\cdot 10^{-3}\,E_0\Rightarrow \xi\equiv\hbar v_{\rm F}/\Delta=283\,a_0$, zero phase difference). The two panels are for a weak and strong perpendicular magnetic field, both at a low temperature $k_{\rm B}T/\Delta=10^{-2}$ in the short-junction regime $L/\xi=0.1$. The cyclotron radius $l_{\rm cycl}$ remains large compared to $L$ also for the strongest fields considered, $l_{\rm cycl}/L= (W/a_0)(\Phi_0/\Phi)\gtrsim 10$. A bipartite square lattice of vortex-antivortex pairs in the normal region (lattice constant $a_{\rm vortex}=\pi l_m^2/L$) forms in the lower panel. Notice the edge reconstruction of the vortex lattice, producing an edge channel of width $\simeq l_m$ large compared to $a_{\rm vortex}$. This edge channel results purely from magnetic interference, it is unrelated to the skipping orbits along the edge that would form in higher fields (when $l_{\rm cycl}< L$).
}
\label{fig_vortexlatticesim}
\end{figure}

As shown in Fig.\ \ref{fig_vortexlatticesim} (resulting from a numerical simulation discussed in Sec.\ \ref{numerics}), in the weak-field regime $l_m\gtrsim L$ there is only a single row of $W/a_{\rm vortex}$ vortex-antivortex pairs. However, when $l_m$ drops well below $L$ multiple rows of vortex-antivortex pairs appear. The appearance of this 2D vortex lattice is associated with a crossover from a $1/B$ to a $1/\sqrt{B}$ decay of the amplitude of the Fraunhofer oscillations. In contrast, for a circular Fermi surface the amplitude crosses over to an accelerated $1/B^2$ decay when $l_m<L$.\cite{Mei16}

The outline of this paper is as follows. In Secs.\ \ref{description} and \ref{semiclassics} we formulate the problem of magnetic interference in a ballistic Josephson junction and present the semiclassical analytical solution for the current distribution. The resulting vortex lattice is described in Sec.\ \ref{currentvortices}, far from the lateral boundaries. As shown in Sec.\ \ref{edgechannel}, within a magnetic length $l_m$ from the boundaries there is a lattice reconstruction that produces an edge channel purely as a result of quantum interference, at magnetic fields that are still so weak that the curvature of the trajectories due to the Lorentz force can be neglected. Because of the edge channel the amplitude of the Fraunhofer oscillations decays as $l_m/W\propto B^{-1/2}$ rather than as $l_m^2/LW\propto B^{-1}$, see Sec.\ \ref{highBdecay}. In Sec.\ \ref{numerics} we test the semiclassics with a fully quantum mechanical solution of a tight-binding model. This numerical simulation also allows us to assess the sensitivity of the results against the effects of disorder and nonideal NS interfaces. We conclude in Sec.\ \ref{conclude}.

\section{Description of the problem}
\label{description}

\begin{figure}[tb]
\centerline{\includegraphics[width=0.8\linewidth]{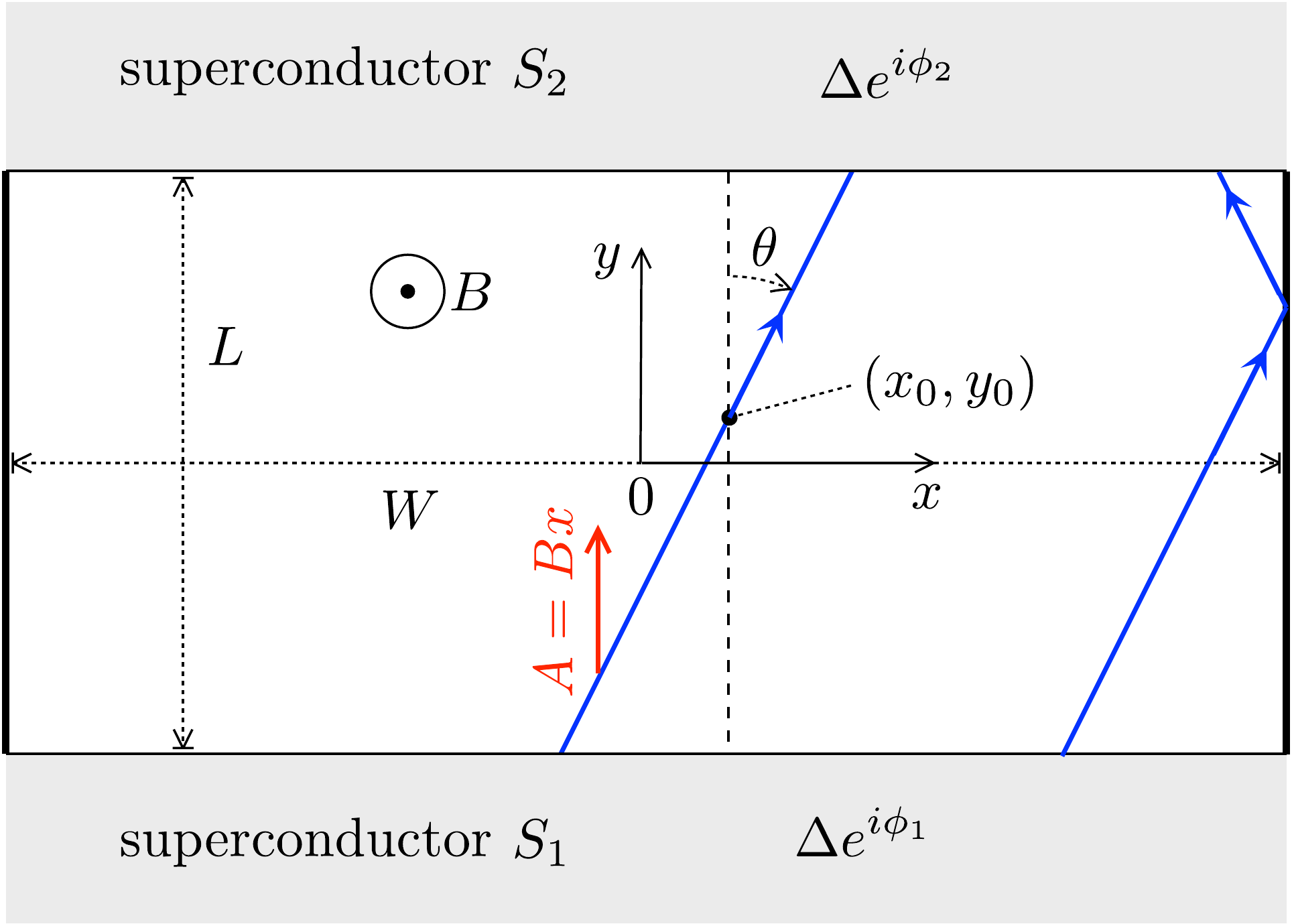}}
\caption{Josephson junction formed by a normal metal (width $W$, length $L$) connecting two superconductors at a phase difference $\phi=\phi_1-\phi_2$. A perpendicular magnetic field $B$ is applied to the normal region. Electron trajectories used in the semiclassical calculation of the supercurrent density are indicated.
}
\label{fig_diagram}
\end{figure}

We consider a two-dimensional (2D) normal metal (N) layer in the $x$--$y$ plane, covered by two superconducting electrodes ($S_1$ and $S_2$) a distance $L$ apart (see Fig.\ \ref{fig_diagram}). The proximity effect induces an excitation gap $\Delta$ in the S-region $|x|<W/2$, $|y|>L/2$, producing a discrete excitation spectrum in the N-region $|x|<W/2$, $|y|<L/2$. 

We work in the short-junction regime $L\ll\xi$, with $\xi=\hbar v_{\rm F}/\Delta$ the superconducting coherence length induced by the proximity effect. (The short-junction regime is chosen for simplicity, we do not expect our qualitative findings to change when $L$ becomes longer than $\xi$.) The lateral width $W$ of the junction is $\gg L$, it may be comparable to $\xi$. The gap $\Delta_0$ in the bulk superconductors is assumed to be much larger than $\Delta$, with a bulk coherence length $\xi_0$ much smaller than $\xi$.

A perpendicular magnetic field $B$ (magnetic length $l_m=\sqrt{\hbar/eB}$) produces oscillations in the critical current of the Josephson junction (Fraunhofer oscillations), periodic with period $\Phi_0=h/2e$ in the enclosed flux $\Phi=BWL$. We assume that the magnetic field is screened from the S-region by a short screening length in the bulk superconductors, even in the high-field regime $l_m\lesssim L$.

In the analytical calculation we take the semiclassical limit $k_{\rm F}L\gg 1$, in which bound states in the junction can be associated with classical trajectories. The junction is ballistic (no impurity scattering), so the trajectories are arcs of cyclotron radius $l_{\rm cycl}=\hbar k_{\rm F}/eB$. We assume that $k_{\rm F}L$ is sufficiently large that the ratio $l_{\rm cycl}/L=k_{\rm F}L\times (l_m/L)^2$ remains $\gg 1$ for the largest fields considered, so we neglect the curvature of the trajectories in the analytical calculation (but not in the numerics). In particular, skipping orbits along the edge play no role in our analysis.

The single-electron dispersion relation $E_{\bm{k}}$ has a nonisotropic dependence on the 2D wave vector $\bm{k}=(k_x,k_y)$, resulting in a nonisotropic distribution of the velocity $\bm{v}_{\bm{k}}=\hbar^{-1}\partial E_{\bm{k}}/\partial\bm{k}$ over the Fermi surface. Our analysis is general, but for a specific example we consider the warping of the Fermi surface on a square lattice (unit lattice constant), with dispersion relation
\begin{equation}
\begin{split}
&E_{\bm{k}}=E_0-\tfrac{1}{2}E_0(\cos k_x+\cos k_y).\\
&\Rightarrow v_{\bm{k}}=\frac{E_0}{2\hbar}(\sin k_x,\sin k_y).
\end{split}
\label{Eksquare lattice}
\end{equation}

The Fermi surface is deformed from a circle to a square as we raise the Fermi energy from the bottom of the band to the band center. For later use we record the relation at the Fermi energy $E_{\rm F}\in(0,E_0)$ between $k_x$ and the angle of incidence $\theta$ on the NS interface:
\begin{equation}
\begin{split}
&\tan\theta=\frac{v_x}{v_y}=\frac{\sin k_x}{\sqrt{1-(\cos k_x+2E_{\rm F}/E_0-2)^2}},\\
&-k_{\rm F}<k_x<k_{\rm F},\;\;k_{\rm F}={\rm arccos}\,(1-2E_{\rm F}/E_0).
\end{split}
\label{thetakxrelation}
\end{equation}

\section{Semiclassical calculation of the supercurrent}
\label{semiclassics}

In semiclassical (WKB) approximation\cite{Bar69} a bound state at energy $|\varepsilon|<\Delta$ corresponds to a periodic classical trajectory that traverses the junction, accumulating a phase shift that is a multiple of $2\pi$. We distinguish two types of periodic trajectories, one in which an \textit{electron} propagates from superconductor $S_1$ to $S_2$, is Andreev reflected as a hole and retraces its path to $S_1$, and another in which a \textit{hole} propagates from $S_1$ to $S_2$ and retraces its path as an electron. The first path is indicated by $\sigma_{\rm eh}=+1$, the second path by $\sigma_{\rm eh}=-1$. 

For a given periodic trajectory the total phase shift is given by
\begin{equation}
\begin{split}
&\phi_{\rm total}=-2\,{\rm arccos}\,(\varepsilon/\Delta)+\sigma_{\rm eh}(\phi-\gamma),\\
&\gamma=\frac{2e}{\hbar}\int_{S_1}^{S_2}\bm{A}\cdot d\bm{l}.
\end{split}
\label{phitotal}
\end{equation}
The $\varepsilon$-dependent term, which has the same sign for $\sigma_{\rm eh}=\pm 1$, is the phase shift accumulated over a penetration depth in the superconductor (in the Andreev approximation\cite{And64} $\Delta\ll E_{\rm F}$). The $\sigma_{\rm eh}$-dependent terms consist of the contribution from the pair potential in $S_1,S_2$ (phase difference $\phi=\phi_1-\phi_2$) and the phase shift $\gamma$ accumulated in the N-region from the vector potential $\bm{A}=(0,Bx,0)$. 

In the short-junction regime $L\ll\xi$ we may neglect the phase shift in N arising from the energy difference $2\varepsilon$ of electron and hole.\cite{note1} For $0<\phi-\gamma<\pi$ the (spin degenerate) bound state corresponding to this periodic trajectory is at energy $\sigma_{\rm eh}\varepsilon$ with
\begin{equation}
\varepsilon=\Delta \cos(\phi/2-\gamma/2).\label{boundstate}
\end{equation}

A tube of width of the order of the Fermi wave length, extending along the trajectory that passes through the point $(x_0,y_0)$ at an angle $\theta$ with the $y$-axis, can be thought of as a single-mode wave guide connecting the two superconductors. In thermal equilibrium at temperature $T$ the single-mode supercurrent is given by\cite{Bee91}
\begin{align}
&\delta I(x_0,y_0,\theta)=-\tanh\left(\frac{\varepsilon}{2k_{\rm B}T}\right)\frac{2e}{\hbar}\frac{d\varepsilon}{d\phi}\nonumber\\
&=\frac{e\Delta}{\hbar}\sin(\phi/2-\gamma/2)\tanh\left(\frac{\Delta\cos(\phi/2-\gamma/2)}{2k_{\rm B}T}\right),\label{deltajresult}
\end{align}
including a factor of two from the spin degeneracy. The trajectory dependence enters via the phase shift $\gamma\equiv\gamma(x_0,y_0,\theta)$. Notice that, notwithstanding the appearance of the half-phases $\phi/2$, the supercurrent is $2\pi$-periodic in $\phi$ --- as it should be.

The total supercurrent $I$ through the Josephson junction follows upon integration of Eq.\ \eqref{deltajresult} over the phase space of the propagating modes at the Fermi level, with measure $dx_0 dk_x/2\pi$:
\begin{equation}
I=\int \frac{dk_x}{2\pi}\int dx_0\, \delta I(x_0,y_0,\theta_{\bm k}).\label{Itotal}
\end{equation}
There is no dependence of $I$ on $y_0$ because of current conservation. 

In zero magnetic field $B=0\Rightarrow\gamma=0$ the dependence of $\delta I$ on $x_0,y_0,\theta$ disappears, so we recover the familiar expression\cite{Kul77}
\begin{equation}
I_0=k_{\rm F}W\frac{e\Delta}{\pi\hbar}\sin(\phi/2)\tanh\left(\frac{\Delta\cos(\phi/2)}{2k_{\rm B}T}\right)\label{I0def}
\end{equation}
for the supercurrent in a ballistic Josephson junction. The zero-temperature critical current, reached at $\phi=\pi-0^{+}$, is
\begin{equation}
I_{c,0}=k_{\rm F}W\frac{e\Delta}{\pi\hbar}.\label{Ic0def}
\end{equation}

We also require the spatial distribution of the supercurrent density. To avoid notational complexity we assume that there is a one-to-one relation between $k_x\in(-k_{\rm F},k_{\rm F})$ and $\theta_{\bm k}\in(-\pi/2,\pi/2)$. This applies to a warping of the Fermi circle that keeps it singly-connected and convex. For a circular Fermi surface the measure $dk_x\mapsto k_{\rm F}\cos\theta\,d\theta$. Upon warping we have instead
\begin{equation}
\frac{dk_x}{2\pi}\mapsto \frac{k_{\rm F}}{2\pi}\rho(\theta)\cos\theta \,d\theta,\label{Pthetadef}
\end{equation}
with a nonuniform angular profile $\rho(\theta)$. The current density can then be written as
\begin{equation}
\begin{pmatrix}
j_x\\
j_y
\end{pmatrix}=\frac{k_{\rm F}}{2\pi}\int_{-\pi/2}^{\pi/2}d\theta\,\rho(\theta)\begin{pmatrix}
\sin\theta\\
\cos\theta
\end{pmatrix}\delta I(x_0,y_0,\theta),\label{jxytheta}
\end{equation}
with $(\sin\theta,\cos\theta)$ a unit vector in the direction of motion (note that $\theta$ is the angle with the $y$-axis, see Fig.\ \ref{fig_diagram}). This is an intuitive expression, but for the calculations it is more convenient to return to $k_x$ as integration variable,
\begin{equation}
\begin{split}
&j_x(x_0,y_0)=\int \frac{dk_x}{2\pi}\, \delta I(x_0,y_0,\theta_{\bm k})\tan\theta_{\bm k},\\
&j_y(x_0,y_0)=\int \frac{dk_x}{2\pi}\, \delta I(x_0,y_0,\theta_{\bm k}).
\end{split}
\label{jxydensity}
\end{equation}

\section{Supercurrent vortex lattice}
\label{currentvortices}

\begin{figure*}[tb]
\centerline{\includegraphics[width=0.8\linewidth]{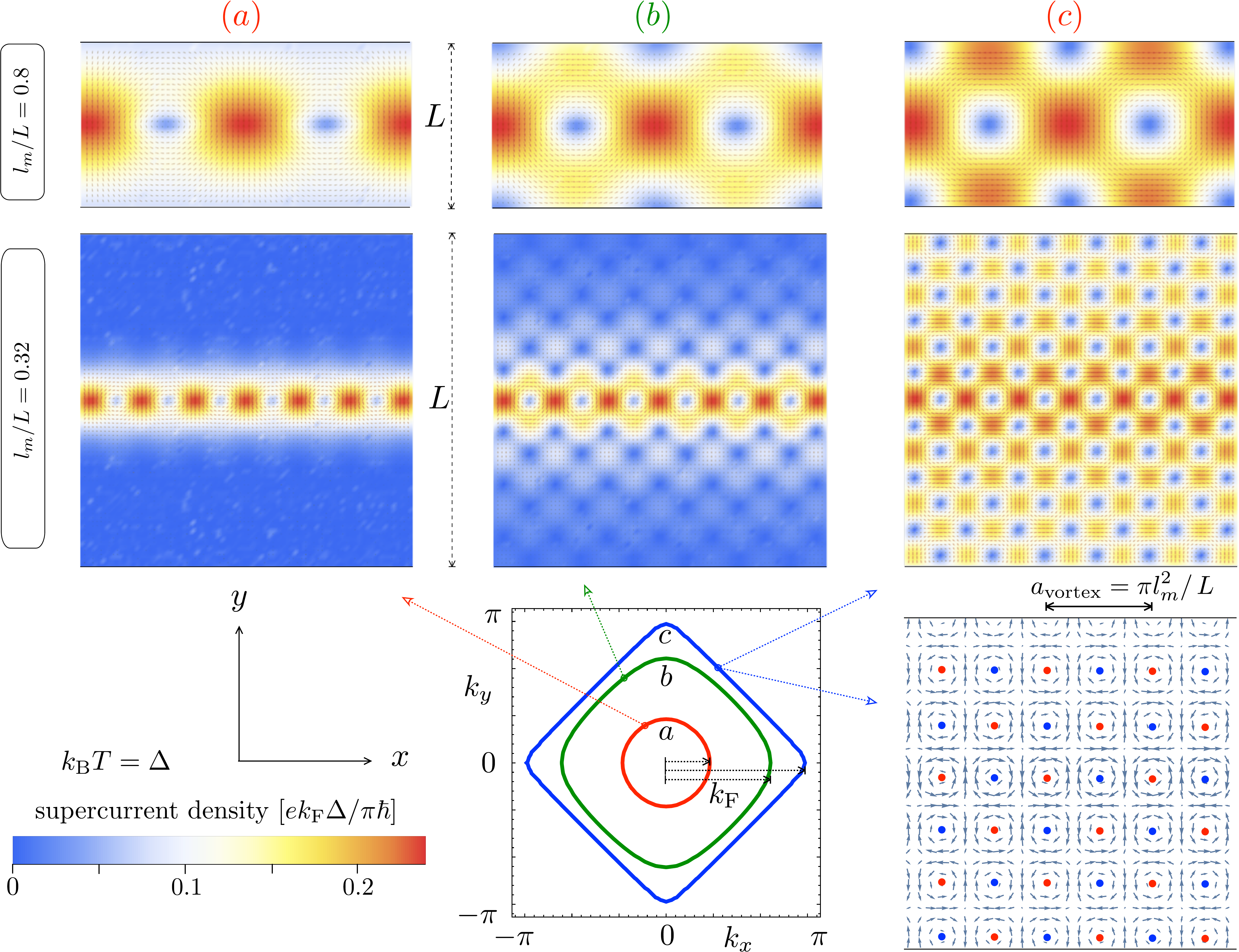}}
\caption{The six color scale plots show the supercurrent density in a wide Josephson junction, far from the lateral boundaries, for two values of the magnetic field (first and second row of panels at $l_m/L=0.8$ and $0.32$, respectively) and for three values of the Fermi energy (labeled $a,b,c$ and corresponding to the square-lattice Fermi surfaces at $E_{\rm F}/E_0=0.2$, $0.8$, and $0.99$, respectively). The plots are calculated from Eqs.\ \eqref{thetakxrelation}, \eqref{jxydensity}, \eqref{gammawide}, at temperature $k_{\rm B}T=\Delta$. The bottom right panel shows the bipartite vortex lattice (vortices and antivortices indicated by red and blue dots, lattice constant $a_{\rm vortex}=\pi l_m^2/L=0.32\,L$ at $l_m/L=0.32$) that develops for $l_m\lesssim L$ in a square-warped Fermi surface.
}
\label{fig_infinitewide}
\end{figure*}

To demonstrate the emergence of a supercurrent vortex lattice we calculate the current density at a point $(x_0,y_0)$ in the normal region, in the limit $W\rightarrow\infty$ that boundary effects can be ignored. (These are considered in the next section.) At a given angle $\theta$ with the $y$-axis (see Fig.\ \ref{fig_diagram}), the phase shift $\gamma$ in Eq.\ \eqref{phitotal} equals
\begin{equation}
\gamma=\frac{2L}{l_m^2}(x_0-y_0\tan\theta).\label{gammawide}
\end{equation}
The resulting current density follows from Eq.\ \eqref{jxydensity} upon integration, once we have specified the relation between $k_x$ and $\theta$. To be definite we take a square lattice dispersion, where $\tan\theta$ is given as a function of $k_x$ by Eq.\ \eqref{thetakxrelation}. Results are shown in Fig.\ \ref{fig_infinitewide}.

If the angular distribution $\rho(\theta)$ on the Fermi surface is peaked at angles $\pm\theta_0$, the phase shift \eqref{gammawide} produces a bipartite rectangular lattice of vortex-antivortex pairs. (Notice that the superconducting phase difference $\phi$ simply shifts the lattice in the $x$-direction.) The lattice constants are $a_\parallel=a_{\rm vortex}$ parallel to the NS interfaces and $a_{\perp}=a_{\rm vortex}/\tan\theta_0$ in the perpendicular direction, with $a_{\rm vortex}$ given by Eq.\ \eqref{avortexdef}.

In the square lattice the Fermi surface has a square warping near the center of the band, and if the NS interfaces are oriented along a principal axis one has $\tan\theta_0=1$, so the vortex-antivortex lattice is a square lattice with lattice constant $a_{\rm vortex}$ in both directions, see panels $(c)$ in Fig.\ \ref{fig_infinitewide}. The two-dimensional lattice disappears --- leaving only a single row of vortices --- if we move away from band center, see panels (a), as the angular distribution $\rho(\theta)$ broadens around normal incidence. Since $a_{\perp}\rightarrow\infty$ for $\theta\rightarrow 0$ this broadening of $\rho(\theta)$ produces a broad range of perpendicular lattice constants, which smear out the structure of the vortex lattice in the direction perpendicular to the NS interface. Only the $\theta$-independent structure parallel to the NS interfaces remains.

\begin{figure}[tb]
\centerline{\includegraphics[width=1\linewidth]{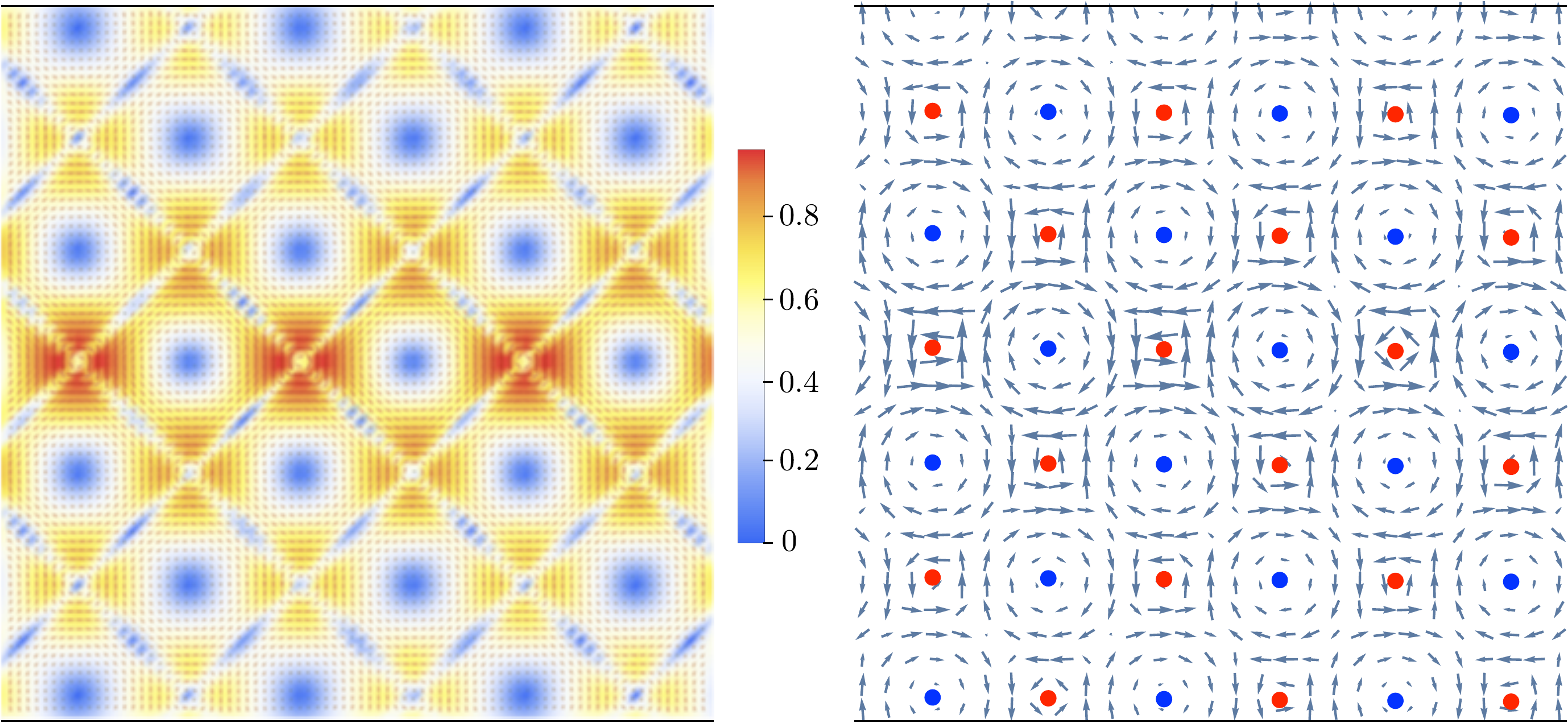}}
\caption{Same as Fig.\ \ref{fig_infinitewide}c for $l_m/L=0.32$, at a much lower temperature of $k_{\rm B}T=0.05\,\Delta$. The vortex and antivortex sublattices (red and blue dots) are no longer equivalent.
}
\label{fig_vortex_lowT}
\end{figure}

At the elevated temperatures $k_{\rm B}T\gtrsim\Delta$ of Fig.\ \ref{fig_infinitewide} the vortices and antivortices are equivalent, but at lower temperatures this symmetry between the two sublattices is broken, see Fig.\ \ref{fig_vortex_lowT}.  Counterclockwise vortices and clockwise antivortices are centered at points where $\phi-\gamma$ equals, respectively, $\pi$ or $0$, modulo $2\pi$. At elevated temperatures the current-phase relationship \eqref{deltajresult} is nearly sinusoidal, with the same slope at $\phi=0,\pi$ (up to a sign difference). At low temperatures the slope at $\phi=0$ is not much affected, so the antivortices retain their circular shape, but the vortices at $\phi=\pi$ see a much larger slope and contract in a square-like shape around the lattice points.

\section{Edge reconstruction of the vortex lattice}
\label{edgechannel}

The vortex lattice is modified if we approach the lateral boundaries at $x=\pm W/2$. We still assume $W\gg L$, so we can treat the  boundaries separately. At each boundary we impose a hard-wall confinement with specular reflection (see Fig.\ \ref{fig_diagram}).

A trajectory from superconductor $S_1$ to $S_2$ that passes through the point $(x_0,y_0)$ at an angle $\theta$ with the $y$-axis is affected by the boundary at $x=W/2$ if $x_0$ is in the interval
\begin{equation}
\tfrac{1}{2}W-\tfrac{1}{2}L|\tan\theta|+y_0\tan\theta<x_0<\tfrac{1}{2}W.\label{x0conditiononecollision}
\end{equation}
In this interval the boundary reflection replaces the expression \eqref{gammawide} for the phase shift $\gamma$ by
\begin{subequations}
\label{gammabetadef}
\begin{align}
\gamma={}&\beta-\frac{1}{2l_m^2|\tan\theta|}(W-2x_0+2y_0\tan\theta)^2,\label{gammaonecollision}\\
\beta={}&\frac{LW}{l_m^{2}}\left(1-\frac{L|\tan\theta|}{2W}\right),\label{betadef}
\end{align}
\end{subequations}
see App.\ \ref{AppABphase}. The corresponding expression for the boundary at $x=-W/2$ follows from the symmetry relation
\begin{equation}
\gamma(x_0,y_0,\theta)=-\gamma(-x_0,y_0,-\theta).\label{gammasymmetry}
\end{equation}

\begin{figure*}[tb]
\centerline{\includegraphics[width=0.75\linewidth]{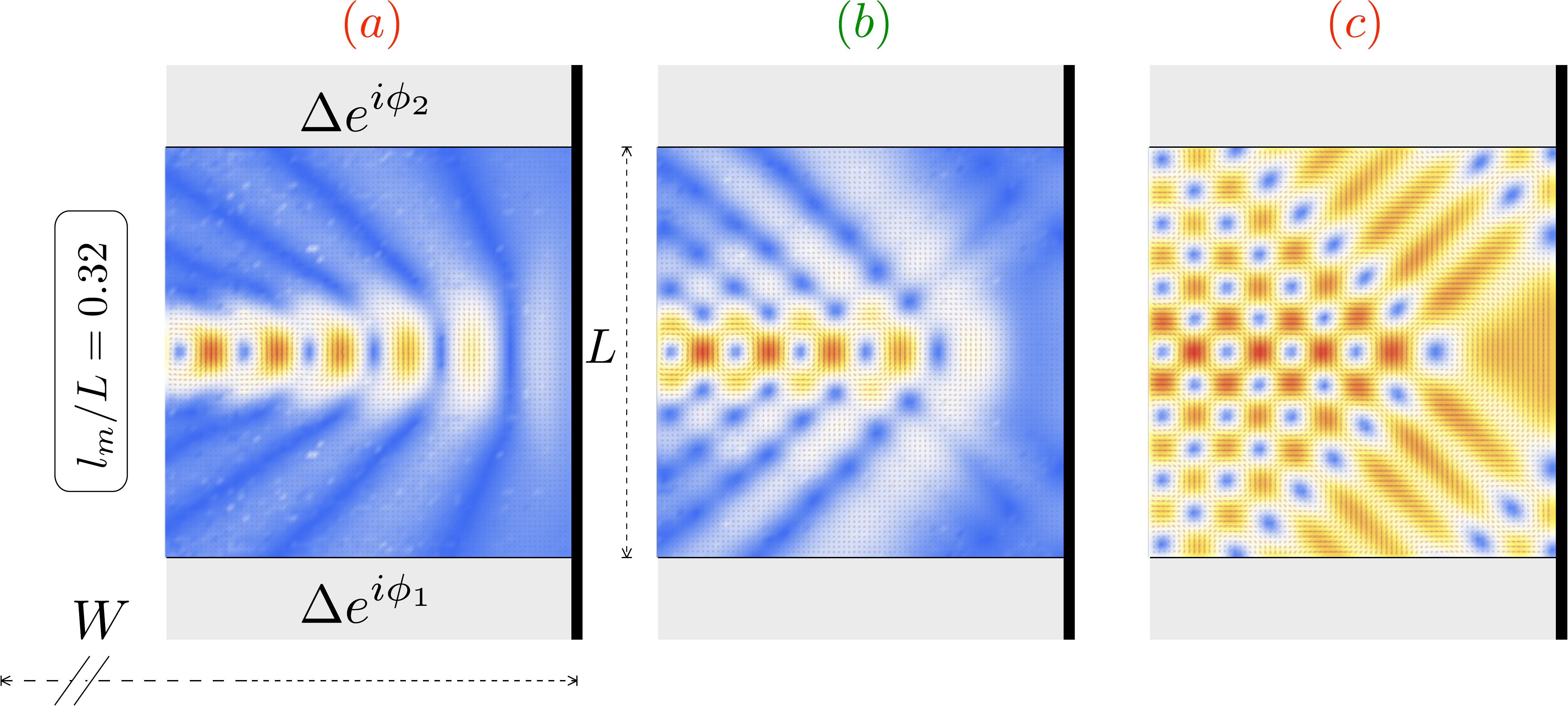}}
\caption{Effect of a hard-wall lateral boundary on the supercurrent vortex lattice . The panels a,b,c correspond to the three labeled Fermi surfaces in Fig.\ \ref{fig_infinitewide}, with the same color scale; the other parameters are $l_m/L=0.32$, $W/L=10.16$, $\phi_1-\phi_2\equiv\phi=\pi/2$, and $k_{\rm B}T=\Delta$.
}
\label{fig_finitewidth}
\end{figure*}

\begin{figure}[tb]
\centerline{\includegraphics[width=0.9\linewidth]{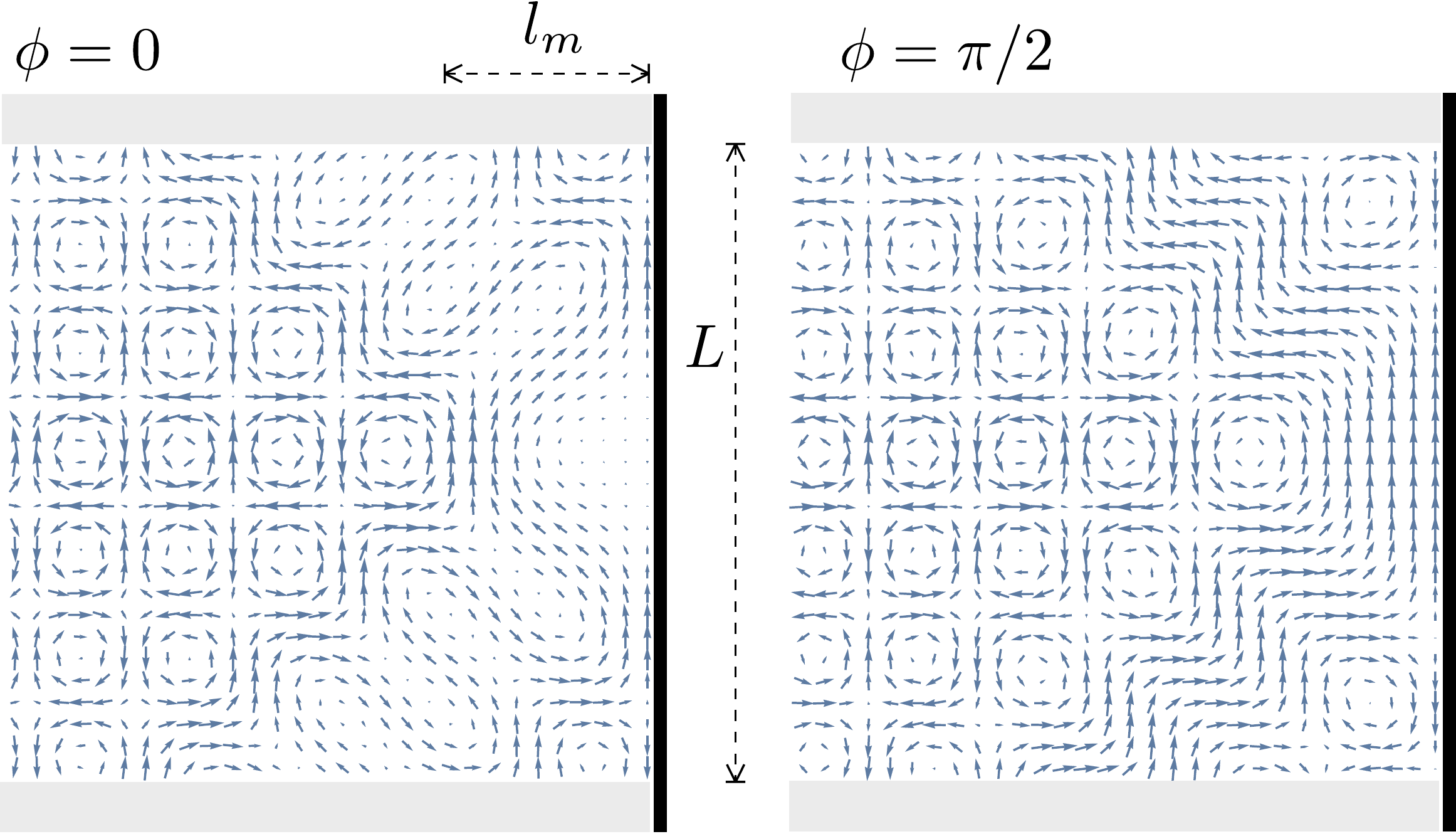}}
\caption{Streamlines corresponding to the vortex lattice in panel (c) of Fig.\ \ref{fig_finitewidth}, for two values of the superconducting phase difference $\phi=\phi_1-\phi_2$ (all other parameters are kept the same). The left and right panels correspond, respectively, to minimal and maximal current flowing along the edge channel.
}
\label{fig_edgevector}
\end{figure}

The resulting supercurrent distribution near the boundary is shown in Fig.\ \ref{fig_finitewidth}. For $l_m\lesssim L$ an edge channel appears when the Fermi surface is strongly warped, see panel (c), becoming less pronounced as the Fermi surface becomes more and more circular, see panels (b) and (a). The streamlines in the edge channel inherit their periodicity from the vortex lattice, but the width $w_{\rm edge}\simeq l_m$ of the edge channel is larger than $a_{\rm vortex}\simeq l_m^2/L$. The net current flowing along the edge channel is sensitive to the phase difference $\phi$ between superconductors $S_1$ and $S_2$, see Fig.\ \ref{fig_edgevector}.

To understand this edge reconstruction of the vortex lattice, we note that because the phase shift $\gamma$ now depends quadratically rather than linearly on $x_0$, there is a point of stationary phase: $\partial\gamma/\partial x_0=0$ at $x_0=y_0\tan\theta+W/2$. For a warped Fermi surface with $\rho(\theta)$ peaked at $\pm\theta_0$ an edge channel extends along the lines of stationary phase, of width
\begin{equation}
w_{\rm edge}\equiv 2\left|\partial^2\gamma/\partial x_0^2\right|^{-1/2}= l_m\sqrt{\tan\theta_0}.\label{wedgedef}
\end{equation}
The edge channel carries a net current from $S_1$ to $S_2$ that depends on the parameter $\beta$ and the superconductor phase difference $\phi$: The edge current is minimal for $\phi-\beta=0$ and maximal for $\phi-\beta=\pi/2$, modulo $\pi$. (In Fig.\ \ref{fig_edgevector} we have $\beta\approx 0$ mod $\pi$, so minimal and maximal current corresponds to $\phi=0$ and $\pi/2$, respectively.) As we will show in the next section, this edge current produces a critical current of order $(w_{\rm edge}/W)I_{c,0}$, with the anomalously slow decay $\propto 1/\sqrt{B}$.

\section{High-field decay of the Fraunhofer oscillations}
\label{highBdecay}

To obtain the critical current $I_c={\rm max}_{\phi}\,I(\phi)$ of the Josephson junction, we first need to calculate at a given phase difference $\phi$ the total supercurrent $I(\phi)$ by integrating $j_y(x_0,y_0)$ over $x_0$ from $-W/2$ to $W/2$. From Eq.\ \eqref{jxytheta} we thus have
\begin{equation}
I=\frac{k_{\rm F}}{2\pi}\int_{-\pi/2}^{\pi/2}\rho(\theta)\cos\theta \,d\theta\int_{-W/2}^{W/2}dx_0\,\delta I(x_0,y_0,\theta).\label{Ictheta}
\end{equation}

Analytical progress is simplest in the high-temperature regime $k_{\rm B}T\gtrsim\Delta$, when the $\phi$-dependence of $\delta I$ from Eq.\ \eqref{deltajresult} becomes approximately sinusoidal,
\begin{equation}
\delta I\approx\frac{e\Delta^2}{4\hbar k_{\rm B}T}\sin(\phi-\gamma),\;\;\gamma=\frac{2e}{\hbar}\int_{S_1}^{S_2}\bm{A}\cdot d\bm{l}.
\label{deltajhighT}
\end{equation}
We assume that the velocity distribution on the Fermi surface is symmetric around normal incidence, $\rho(\theta)=\rho(-\theta)$. Because of Eq.\ \eqref{gammasymmetry} we may then restrict the $\theta$-integration in Eq.\ \eqref{Ictheta} to positive angles,
\begin{align}
I={}&\frac{e\Delta^2 k_{\rm F}}{8\pi\hbar k_{\rm B}T}\int_{0}^{\pi/2}\rho(\theta)\cos\theta\, d\theta\int_{-W/2}^{W/2}dx_0\nonumber\\
&\times[\sin(\phi-\gamma)+\sin(\phi+\gamma)].\label{Icthetaplus}
\end{align}
We thus find that the integrated supercurrent retains a sinusoidal $\phi$-dependence, with critical current
\begin{equation}
\begin{split}
&I_c={}I_{c,0}\left|\int_{0}^{\pi/2}\rho(\theta)\cos\theta\, d\theta\int_{-W/2}^{W/2}\frac{dx_0}{W}\cos\gamma\right|,\\
&I_{c,0}=\frac{e\Delta^2 k_{\rm F}W}{4\pi\hbar k_{\rm B}T}.
\end{split}
\label{IchighT}
\end{equation}

In the interval $0<\theta<{\rm arctan}\,(W/L)$ there is at most one boundary collision. We restrict ourselves to this interval, because the contributions to $I_c$ near grazing incidence are anyway suppressed exponentially at finite temperature. (All contributions are included in the numerics.) Fixing the arbitrary $y$-coordinate at $y_0=-L/2$, we have from Eqs.\ \eqref{gammawide} and \eqref{gammabetadef} the expression for $\gamma$ that we need:
\begin{subequations}
\label{gammaresult}
\begin{align}
\gamma={}&\frac{2L}{l_m^2}(x_0+\tfrac{1}{2}L\tan\theta)\;\;{\rm if}\;\;x_0+L\tan\theta<W/2,\label{gammaresulta}\\
\gamma={}&\beta-\frac{(W-2x_0-L\tan\theta)^2}{2l_m^2\tan\theta}\;\;{\rm if}\;\;x_0+L\tan\theta>W/2,\label{gammaresultb}
\end{align}
\end{subequations}
with $\beta$ defined in Eq.\ \eqref{betadef}. 

The integral over $x_0$ in Eq.\ \eqref{IchighT} can be carried out analytically:
\begin{align}
&I_c=I_{c,0}\left|\int_{0}^{\pi/2}\rho(\theta)\Gamma(\theta)\cos\theta\, d\theta\right|,\label{IcGamma}\\
&\Gamma(\theta)\equiv\int_{-W/2}^{W/2}\frac{dx_0}{W}\cos\gamma=\frac{l_m^2}{LW}\sin\beta'\label{intcosgammaA}\\
&\quad+(l_m/W)\sqrt{\pi\tan\theta}[F_C(\alpha)\cos\beta+F_S(\alpha)\sin\beta],\nonumber\\
&\alpha=\frac{L\sqrt{\tan\theta}}{l_m\sqrt\pi},\;\;\beta'=\frac{LW}{l_m^2}\left(1-\frac{L}{W}\tan\theta\right).\label{alphabetadef}
\end{align}
The functions $F_C$ and $F_S$ are the Fresnel cosine and sine integrals,
\begin{equation}
F_C(\alpha)=\int_0^\alpha\cos(\tfrac{\pi}{2} t^2)\,dt,\;\;F_S(\alpha)=\int_0^\alpha\sin(\tfrac{\pi}{2} t^2)\,dt.\label{FCFSdef}
\end{equation}
Both $F_C(\alpha)$ and $F_S(\alpha)$ tend to $1/2$ for $\alpha\rightarrow\infty$.

\begin{figure}[tb]
\centerline{\includegraphics[width=0.9\linewidth]{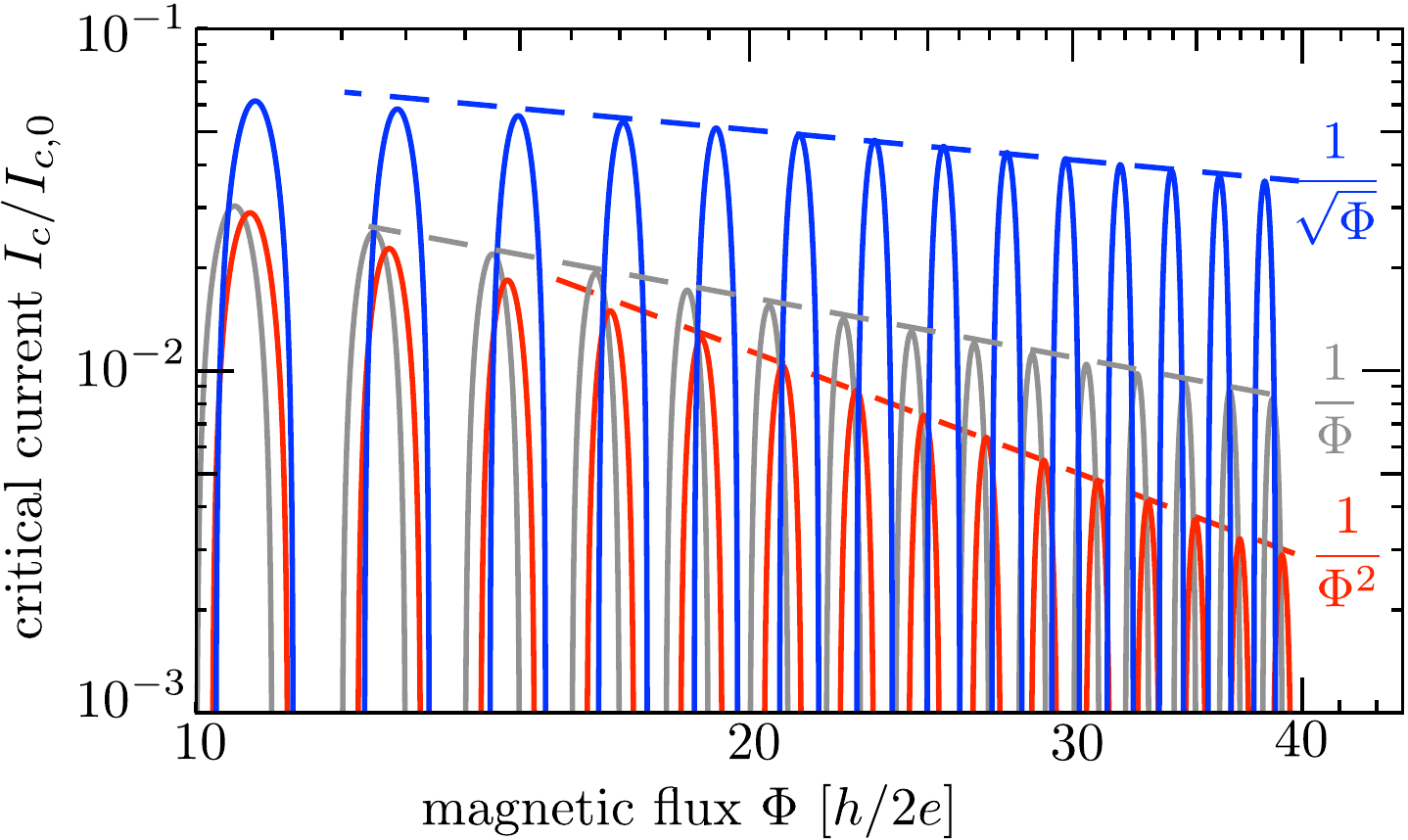}}
\caption{Log-log plot of the critical current $I_c$ versus the flux $\Phi$ through the normal region (aspect ratio $W/L=10.16$), calculated from Eq.\ \eqref{IcGamma} for a circular Fermi surface ($\rho(\theta)=1$, red curve decaying $\propto 1/\Phi^2$), and for a square Fermi surface ($\rho(\theta)=\delta(\theta-\pi/4)$, blue curve decaying $\propto 1/\sqrt\Phi$). The low-field Fraunhofer oscillations \eqref{Iclowfield} are included for comparison (grey curve decaying $\propto 1/\Phi$). 
}
\label{fig_Fraunhofer}
\end{figure}

If the angular distribution $\rho(\theta)$ is sharply peaked around $\pm\theta_0$, we obtain from Eqs.\ \eqref{IcGamma} and \eqref{intcosgammaA} the high-field ($l_m\ll L$) critical current
\begin{equation}
I_c(\text{high-field})=I_{c,0}\frac{w_{\rm edge}}{W}\sqrt{\pi/2}\left|\sin\left(\frac{\pi}{4}+\frac{LW_{\rm eff}}{l_m^2}\right)\right|,\label{Ichighfield}
\end{equation}
with effective junction width $W_{\rm eff}=W-\tfrac{1}{2}L\tan\theta_0$ and edge channel width $w_{\rm edge}=l_m\sqrt{\tan\theta_0}$. Comparing with the low-field ($l_m\gg L$) Fraunhofer oscillations,
\begin{equation}
I_c(\text{low-field})=I_{c,0}\frac{l_m^2}{LW}\left|\sin(LW/l_m^2)\right|,\label{Iclowfield}
\end{equation}
we note three differences: the amplitude decays more slowly, $\propto 1/\sqrt{B}$ instead of $\propto 1/B$; the flux periodicity is larger by a factor $W/W_{\rm eff}$; and the maxima are phase shifted by $1/4$ flux quantum. This qualitatively different behavior is illustrated in Fig.\ \ref{fig_Fraunhofer}, compare blue and grey curves.

At the other extreme of an isotropic angular distribution, for a circular Fermi surface, we obtain the opposite effect: instead of a slower decay of the high-field Fraunhofer oscillations the decay is faster, $\propto 1/B^2$ instead of $\propto 1/B$, compare red and blue curves.\cite{note2} This accelerated decay is a known result.\cite{Mei16} What we have found here is that the switch from a circular to a square Fermi surface slows down the decay by a fourth root, from $B^{-2}$ to $B^{-1/2}$.

\section{Numerical simulations}
\label{numerics}

To test the analytical semiclassical theory we have performed numerical simulations of a tight-binding model. We start from the Bogoliubov-De Gennes Hamiltonian, 
\begin{equation}
H(\bm{k})=\begin{pmatrix}
E(\bm{k}-e\bm{A})-E_{\rm F}&\Delta\\
\Delta^\ast&E_{\rm F}-E(\bm{k}+e\bm{A})
\end{pmatrix},\label{HBdGdef}
\end{equation}
with the single-particle dispersion $E(\bm{k})$ on a square lattice given by Eq.\ \eqref{Eksquare lattice}. The pair potential $\Delta$ and vector potential $\bm{A}$ are chosen as in Fig.\ \ref{fig_diagram}, with $\Delta=0$ for $|y|<L/2$ (no pairing interaction in the normal region) and $\bm{A}=0$ for $|y|>L/2$ (complete screening of the magnetic field from the superconductor). The self-field of the currents in the normal region is neglected, so $\bm{A}$ is entirely due to the externally imposed field $B$. The orbital effect of the magnetic field is fully included, but we neglect the coupling to the electron spin\cite{note4} and can therefore omit the spin degree of freedom from the Hamiltonian.

The $2\times 2$ matrix Green's function $G(\varepsilon)=(\varepsilon-H)^{-1}$ is calculated at imaginary energy $\varepsilon=i\omega$ using the Kwant toolbox for tight-binding models.\cite{kwant} The expectation value of the current density in thermal equilibrium,
\begin{equation}
\bm{j}(\bm{r})=\frac{2e}{\hbar}k_{\rm B}T\,{\rm Re}\,\sum_{p=0}^\infty {\rm Tr}\,\langle\bm{r}|G(i\omega_p)|\bm{r}\rangle\langle\bm{r}|\frac{\partial H}{\partial\bm{k}}|\bm{r}\rangle,
\end{equation}
is then obtained from a (rapidly convering) sum over Matsubara frequencies $\omega_p=(2p+1)\pi k_{\rm B}T$.\cite{Fur94} (See Ref.\ \onlinecite{Rak15} for an alternative approach.) 

The time-consuming step in this calculation is the calculation of the inverse operator $(i\omega-H)^{-1}$, but once this is done for one value of the superconducting phase difference $\phi$, we can use Dyson's equation to obtain the result for other values of $\phi$ without further inversions. 

\begin{figure}[tb]
\centerline{\includegraphics[width=1\linewidth]{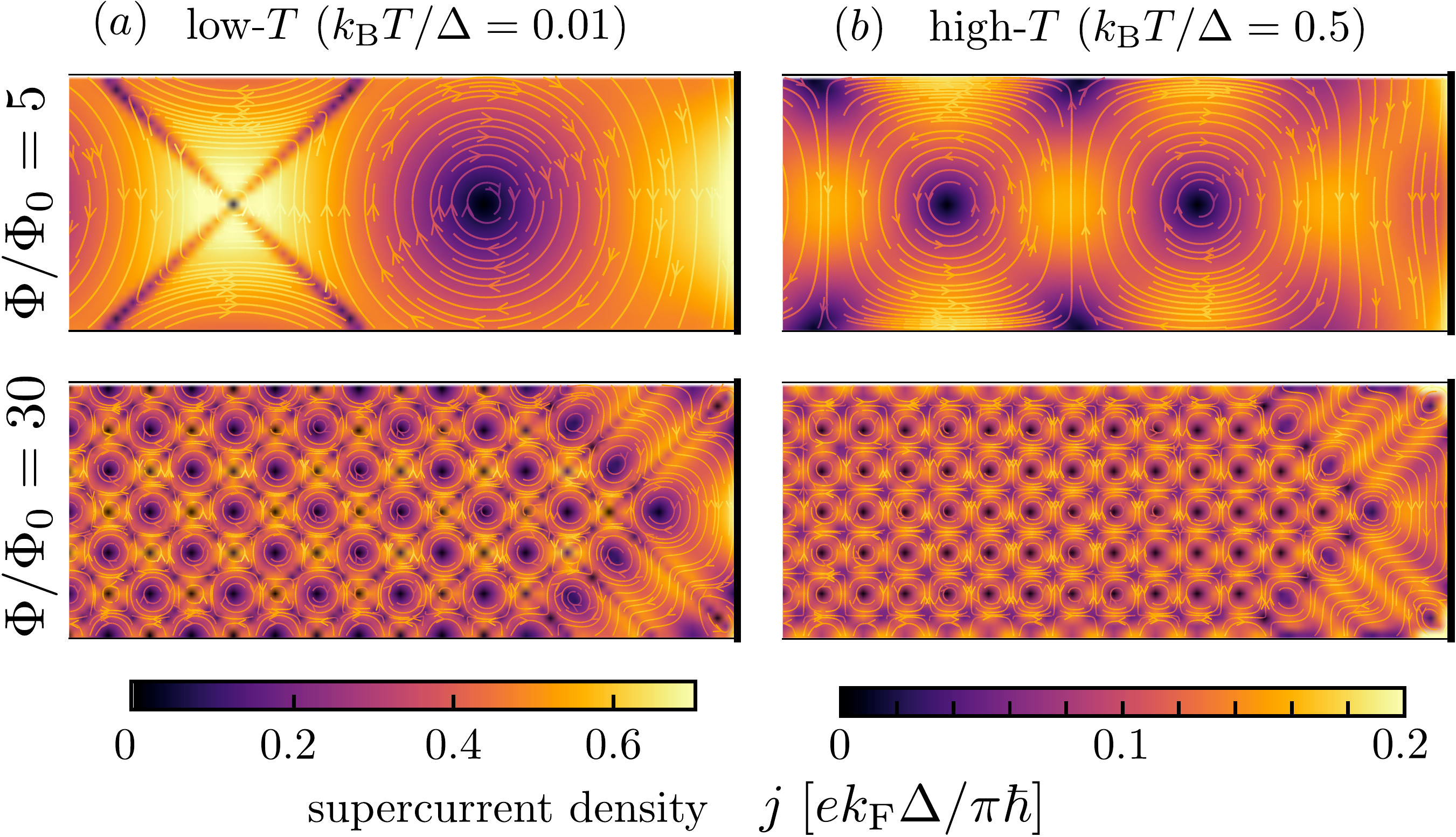}}
\caption{(a): Same as Fig.\ \ref{fig_vortexlatticesim}, zoomed in at the right boundary. (b): At a higher temperature the vortices and antivortices are approximately equivalent.
}
\label{fig_vortexlatticehighT}
\end{figure}

\begin{figure}[tb]
\centerline{\includegraphics[width=0.8\linewidth]{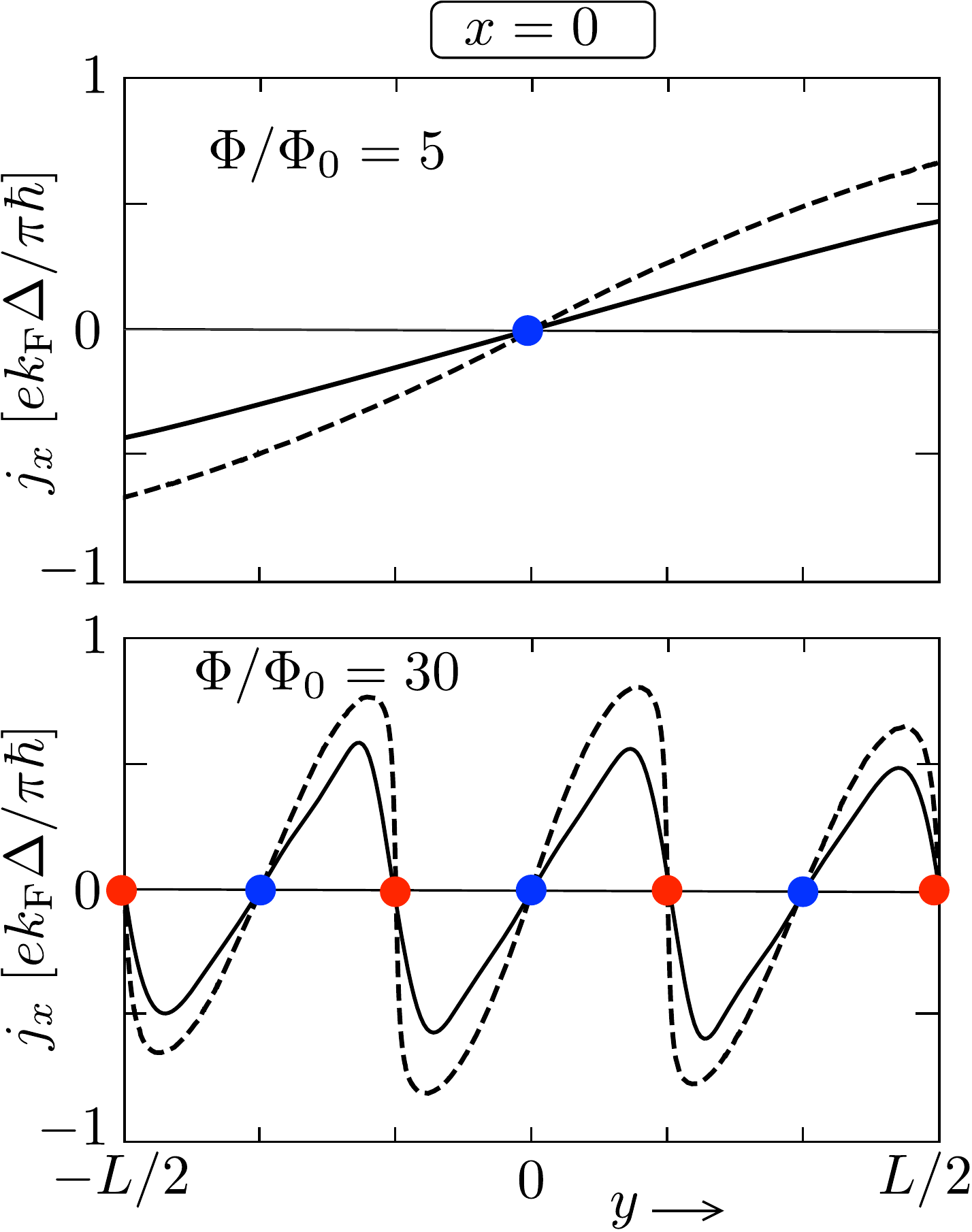}}
\caption{Current density profile along a cut through $x=0$, for the same parameters as Fig.\ \ref{fig_vortexlatticesim}. Since $j_y=0$ along this cut, the plotted $j_x$ is the full current density. The red and blue dots identify the center of a vortex or antivortex, which are distinct at this low temperature of $k_{\rm B}T=0.01\,\Delta$. The solid curves are the results of the numerical simulation, the dashed curves are the semiclassical result \eqref{jxydensity} in the short-junction regime.
}
\label{fig_comparison_lowT_highB}
\end{figure}

\begin{figure}[tb]
\centerline{\includegraphics[width=0.7\linewidth]{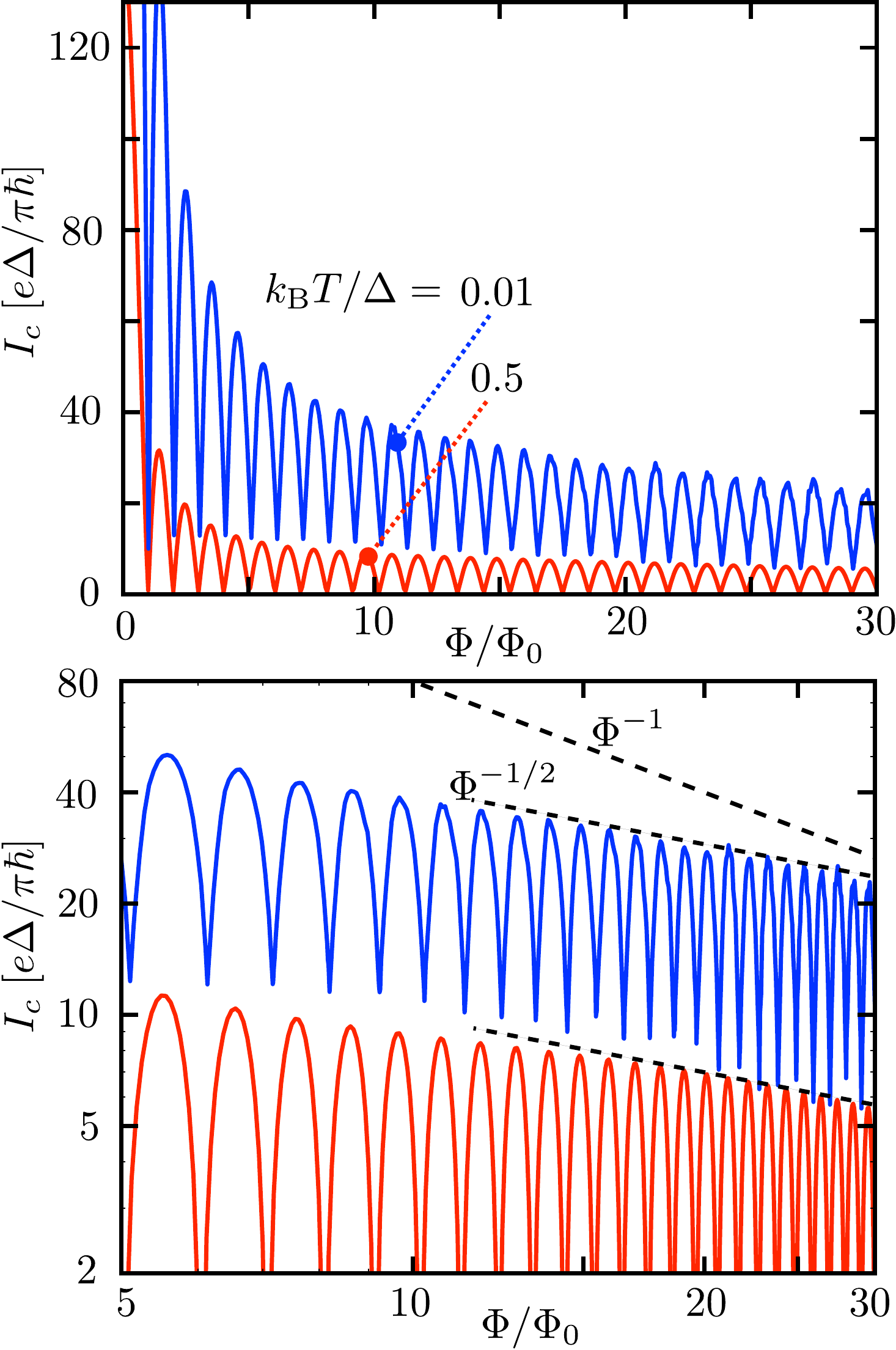}}
\caption{Plot of the critical current $I_c$ versus the flux $\Phi$ through the normal region, resulting from the numerical simulation with the parameters of Figs.\ \ref{fig_vortexlatticesim} and \ref{fig_vortexlatticehighT}. The minima of the Fraunhofer oscillations no longer go to zero at low temperatures (blue curves), because of the skewed current-phase relationship. The upper panel shows a linear scale, the lower panel a log-log scale with the $\Phi^{-1/2}$ decay indicated (black dashed line). (The $1/\Phi$ decay of the conventional Fraunhofer oscillations is also included for comparison.)
}
\label{fig_Fraunhofersim}
\end{figure}

Results for the vortex lattice in the case of a nearly square Fermi surface ($E_{\rm F}/E_0=0.99$) are shown in Figs.\ \ref{fig_vortexlatticesim} and  \ref{fig_vortexlatticehighT}. The agreement with the semiclassical result is not fully quantitative, see Fig.\ \ref{fig_comparison_lowT_highB}, but all the qualitative features of the vortex lattice coming out of the analytics are well reproduced in the numerics. Also the $1/\sqrt B$ decay is recovered in the simulation, see Fig.\ \ref{fig_Fraunhofersim}. 

\begin{figure}[tb]
\centerline{\includegraphics[width=0.8\linewidth]{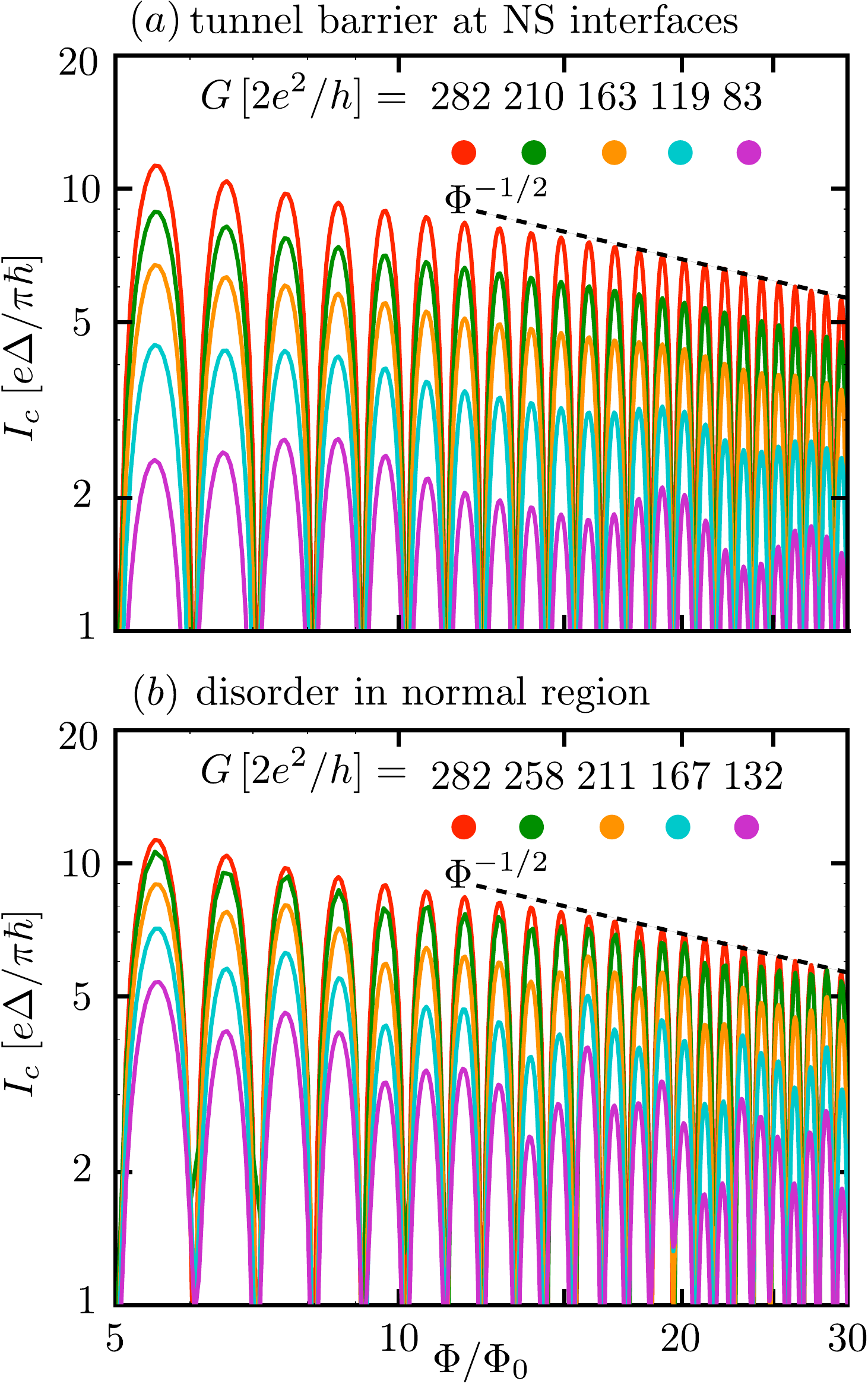}}
\caption{Effect on the Fraunhofer oscillations of a tunnel barrier at the NS interfaces (panel a) or of disorder in the normal region (panel b). The data results from the numerical simulation with the parameters of Fig.\ \ref{fig_vortexlatticehighT}b. The disorder strength or tunnel barrier height is quantified by the reduction of the normal state conductance $G$. The topmost (red) curve corresponds to the ideal case without disorder or tunnel barrier.
}
\label{fig_Fraunhoferdisorder}
\end{figure}

In both the analytics and numerics so far we took a ballistic Josephson junction, without any disorder in the normal region, and ideal (fully transparent) NS interfaces. The numerical simulation provides a way to test for the effects of impurity scattering and nonideal interfaces. Disorder was modeled by adding a random component $\delta U$ to the on-site electrostatic potential, drawn uniformly from the interval $[-U_0,U_0]$. For the tunnel barrier we reduced the hopping amplitude at the two NS interfaces. As shown in Fig.\ \ref{fig_Fraunhoferdisorder}, the slow $1/\sqrt{B}$ decay persists even if the critical current is reduced substantially by the tunnel barrier. Disorder provides a stronger perturbation, in the form of random sample-specific fluctuations,\cite{Mei16} but averaged over series of peaks the slow decay persists. 

\section{Discussion}
\label{conclude}

\textit{Two-dimensional} vortex lattices are well established for Abrikosov vortices in a bulk superconductor,\cite{Tinkham} but Josephson vortices in an SNS junction were only known to arrange as a \textit{one-dimensional} chain.\cite{Cue07,Ber08,Ali15} Our key conceptual finding is that the 2D arrangement is hidden by angular averaging over the Fermi surface. For a distribution of angles of incidence peaked at $\pm\theta$, resulting from a strong square or hexagonal warping of the Fermi surface, a 2D lattice develops when the magnetic length $l_m=\sqrt{\hbar/eB}$ drops below the separation $L$ of the NS interfaces. The lattice is bipartite, with a vortex and antivortex in a rectangular unit cell of size $\pi l_m^2/L$ parallel to the interface and $\pi l_m^2/(L\tan\theta)$ perpendicular to the interface. For a circular Fermi surface the 2D lattice degrades to a 1D chain.

\begin{figure}[tb]
\centerline{\includegraphics[width=1\linewidth]{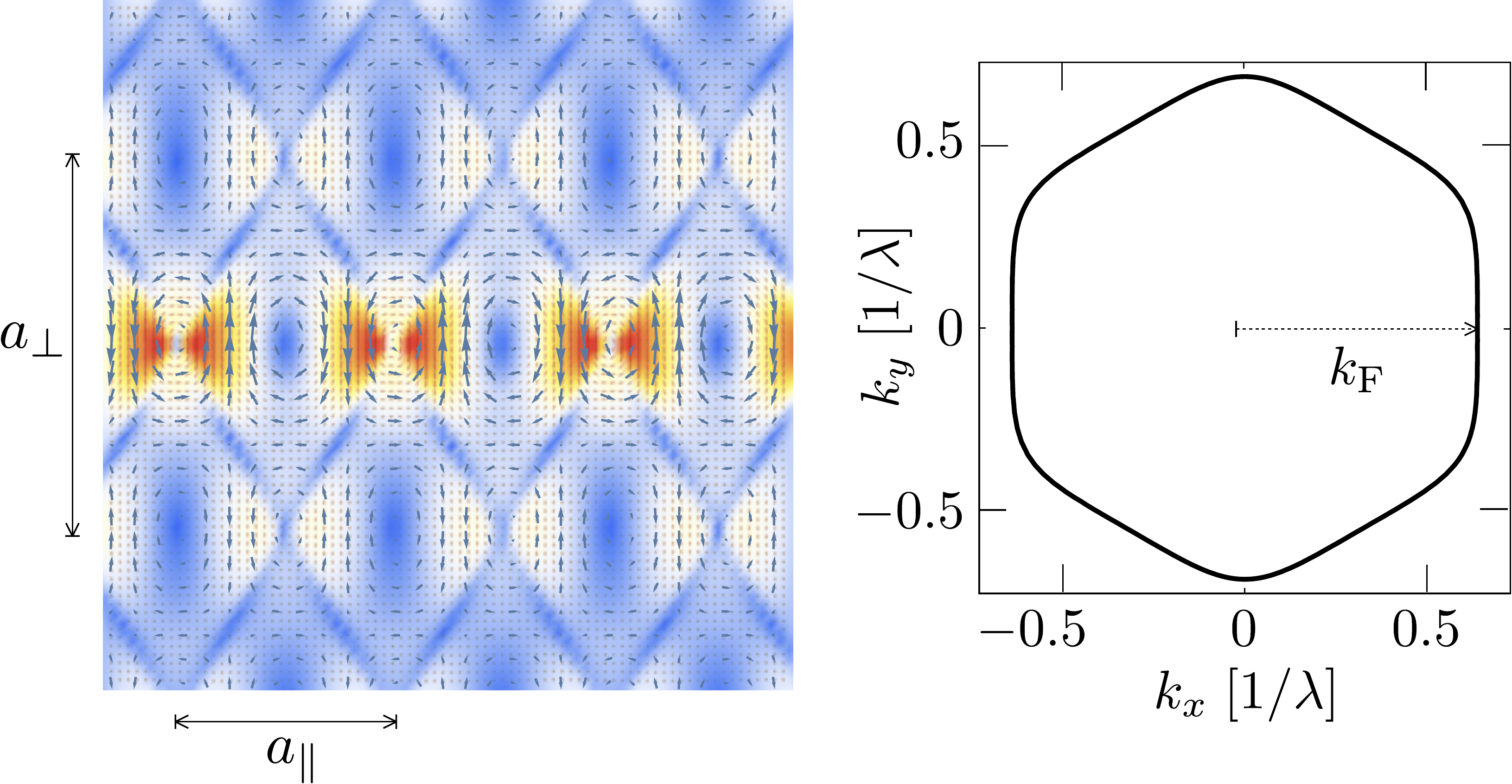}}
\caption{Vortex lattice for a Fermi surface having the hexagonal warping of the ${\rm Bi}_2{\rm Te}_3$ dispersion relation \eqref{EkBi2Te3} (parameters $\lambda\approx 1\,{\rm nm}$, $E_0\approx 260\,{\rm meV}$, $E_{\rm F}=6^{-3/4}\sqrt{7}\,E_0$, $k_{\rm F}=6^{-1/4}\lambda^{-1}$, other parameters and color scale as in Fig.\ \ref{fig_vortex_lowT}). The difference with square warping is that the lattice is rectangular rather than square, with aspect ratio $a_\perp/a_\parallel=1/\tan(\pi/6)=\sqrt 3$.
}
\label{fig_hexagon}
\end{figure}

It would be interesting to search for this 2D Josephson vortex lattice in some of the quasi-two-dimensional systems that are known to have a warped Fermi surface, such as the hexagonal warping on the surface of a three-dimensional topological insulator.\cite{Has09} By way of illustration, Fig.\ \ref{fig_hexagon} shows the vortex lattice calculated for the $[111]$ surface dispersion of ${\rm Bi}_2{\rm Te}_3$,\cite{Fu09}
\begin{equation}
E_{\bm k}=E_0\sqrt{\lambda^2 k_x^2+\lambda^2 k_y^2+\lambda^6(k_x^3-3k_xky^2)^2},\label{EkBi2Te3}
\end{equation}
with the $x$-axis (the NS interface) oriented along the $\Gamma$K direction in the Brillouin zone.

The vortices could be detected directly by a scanning tunneling probe,\cite{Aus09,Fin12,Rod15} or indirectly through the flux $\Phi$-dependent Fraunhofer oscillations\cite{Chi12,Cro13} --- we have found that the transition from a 1D to a 2D arrangement of vortices is accompanied by a slow-down of the decay of the oscillation amplitude from $1/\Phi$ to $1/\sqrt{\Phi}$. While in the main text we have focused on the current distribution, we note that a 2D lattice structure with the same periodicity appears also in the superconducting pair potential (see App.\ \ref{Apporderparam}) and in the local density of states (see App.\ \ref{AppDOS}).

A particularly intriguing feature of the vortex lattice is the reconstruction at the edge, resulting in an edge channel of width $\simeq l_m$ parametrically larger than the lattice constant. It is this edge channel that effectively carries the supercurrent when $l_m\lesssim L$, resulting in the decay scaling as $l_m/W\propto 1/\sqrt{B}$. Notice that the edge channel appears entirely as a result of quantum interference --- in contrast to the quantum Hall edge channel any orbital effects of the magnetic field play no role here.

\acknowledgments
We have benefited from discussions with I. Muhammad and M. Wimmer. This research was supported by the Foundation for Fundamental Research on Matter (FOM), the Netherlands Organization for Scientific Research (NWO/OCW), and ERC Starting and Synergy Grants.

\appendix

\section{Calculation of the Aharonov-Bohm phase shift}
\label{AppABphase}

We calculate the Aharonov-Bohm phase shift
\begin{equation}
\gamma=\frac{2e}{\hbar}\int_{S_1}^{S_2}\bm{A}\cdot d\bm{l}\label{gammaABdef}
\end{equation}
accumulated along a trajectory across the Josephson junction, from superconductor $S_1$ at $y=-L/2$ to $S_2$ at $y=+L/2$, including the effects of multiple specular reflections at the side walls $x=\pm W/2$. The geometry is shown in Fig.\ \ref{fig_diagram}. Assume that the trajectory starts at $t=0$ from the point $x=x(0)$, $y=-L/2$ at the lower NS interface, at an angle $\theta(0)\in(-\pi/2,\pi/2)$ with the positive $y$-axis. The opposite NS interface at $y=L/2$ is reached at the time $t_L=L/v_y$, with $v_y=v_{\rm F}\cos\theta(0)$ the velocity component in the $y$-direction (which does not change at a boundary reflection).

In the gauge $\bm{A}=(0,Bx,0)$ the line integral takes the form
\begin{equation}
\gamma=\frac{2v_y}{l_m^2}\int_0^{t_L} x(t)dt.\label{gammatLdef}
\end{equation}
The time dependence of $x(t)$ is given by
\begin{equation}
\begin{split}
&x(t)=(-1)^{\nu_{u(t)}}[u(t)-\nu_{u(t)}W],\\
&u(t)=x(0)+v_{\rm F}t\sin\theta(0),
\end{split}
\end{equation}
where we have defined $\nu_{u}\in\mathbb{Z}$ as the integer nearest to $u/W$. The absolute value of $\nu$ counts the number of boundary reflections up to time $t$. At time $t_L=L/[v_{\rm F}\cos\theta(0)]$ we have
\begin{equation}
x(t_L)=(-1)^{\nu_L}[x(0)+L\tan\theta(0)-\nu_L W],
\end{equation}
where $\nu_L\equiv\nu_{u(t_L)}$ is the integer nearest to $[x(0)+L\tan\theta(0)]/W$.

Integration of Eq. \eqref{gammatLdef} results in
\begin{equation}
\gamma=\frac{1}{l_{m}^2\tan\theta(0)}\,\biggl(\tfrac{1}{4}W^2-x^2(0)+(-1)^{\nu_L} \bigl[x^2(t_L)-\tfrac{1}{4}W^2\bigr]\biggr).\label{gammaApp}
\end{equation}
This is sufficient to calculate the total current through the Josephson junction, by integrating the current density through the lower NS interface.

To obtain the current distribution within the junction, say at the point $(x_0,y_0)$, we need to find the corresponding coordinates $(x(0),-L/2)$ of the trajectory at the lower NS interface. The angle $\theta$ at the point $(x_0,y_0)$ equals $\pm\theta(0)$. The point $(x_0,y_0)$ is reached at a time $t_0=(y_0+L/2)/v_y$ after
\begin{equation}
\nu_0=\nu_{x_0-v_{\rm F}t_0\sin\theta}=\nu_{x_0-(y_0+L/2)\tan\theta}
\end{equation}
boundary reflections. Retracing back the trajectory, we find
\begin{equation}
\begin{split}
&x(0)=(-1)^{\nu_0}[x_0-(y_0+L/2)\tan\theta-\nu_0 W],\\
&\theta(0)=(-1)^{\nu_0}\theta.
\end{split}
\end{equation}

This calculation of the Aharonov-Bohm phase $\gamma$ holds for any number of boundary collisions at $x=\pm W/2$. In the main text we only need the result for a single boundary collision at $x=W/2$. One readily checks that Eq.\ \eqref{gammaApp} reduces to Eq.\ \eqref{gammabetadef} upon substitution of $\nu_L=1$, $\nu_0=0$ for $\tan\theta>0$ or $\nu_L=1$, $\nu_0=1$ for $\tan\theta<0$.

\section{Two-dimensional lattice structure of the superconducting order parameter}
\label{Apporderparam}

The coherent superposition of electrons and holes in an Andreev level produces a nonzero order parameter $F(\bm{r})$ in the normal region, in the absence of any pairing interaction.\cite{Tinkham} In this appendix we show that the amplitude $|F|$ has a 2D lattice structure with the same periodicity as the current vortex lattice studied in the main text.

An Andreev level in the SNS junction of Fig.\ \ref{fig_diagram}, at the positive energy 
\begin{equation}
\varepsilon=\Delta \cos(\psi/2),\;\;\psi=\phi_1-\phi_2-\gamma\in(-\pi,\pi),
\end{equation}
has a wave function $\Psi(\bm{r})$ that penetrates into the superconducting regions $|y|>L/2$ over a distance
\begin{equation}
\xi_\varepsilon=\hbar v_y (\Delta^2-\varepsilon^2)^{-1/2}=(\hbar v_y/\Delta)|\sin(\psi/2)|^{-1}.
\end{equation}
In the normal region $|y|<L/2$ the wave function has a constant amplitude, given in WKB approximation by\cite{Bar69}
\begin{equation}
\Psi(\bm{r})=\begin{pmatrix}
u(\bm{r})\\
v(\bm{r})
\end{pmatrix}=(2\xi_\varepsilon)^{-1/2}e^{i\bm{k}\cdot\bm{r}}
\begin{pmatrix}
e^{i\eta/2}\\
e^{-i\eta/2}
\end{pmatrix}.
\end{equation}
The electron and hole components $u,v$ differ in phase by
\begin{equation}
\eta=\tfrac{1}{2}(\phi_1+\phi_2+\gamma)-\frac{2e}{\hbar}\int_{S_1}^{\bm{r}} \bm{A}\cdot d\bm{l},\label{etadef}
\end{equation}
in accord with the Andreev reflection boundary condition at the NS interfaces,\cite{Bee91}
\begin{equation}
\eta=\begin{cases}
\phi_1-\sigma\,{\rm arccos}\,(\varepsilon/\Delta)&\text{at}\, y=-L/2,\\
\phi_2+\sigma\,{\rm arccos}\,(\varepsilon/\Delta)&\text{at}\, y=+L/2.
\end{cases}
\end{equation}
We have defined $\sigma={\rm sign}\,\psi$, so that ${\rm arccos}\,(\varepsilon/\Delta)=\sigma\psi/2$ for $\psi\in(-\pi,\pi)$.

The electron-hole mode $(u,v)$ at energy $\varepsilon$ contributes to the superconducting order parameter an amount\cite{Tinkham}
\begin{equation}
\delta F(\bm{r})=\tanh\left(\frac{\varepsilon}{2k_{\rm B}T}\right)u^\ast(\bm{r}) v^{\vphantom{\ast}}(\bm{r}).
\end{equation}
Integration over the modes gives the full order parameter,
\begin{align}
&F(\bm{r})=\int \frac{dk_x}{2\pi}\delta F(\bm{r})\nonumber\\
\quad &=\frac{k_{\rm F}}{2\pi}\int_{-\pi/2}^{\pi/2} d\theta\,\rho(\theta)\cos\theta \tanh\left(\frac{\varepsilon}{2k_{\rm B}T}\right)\frac{e^{-i\eta}}{2\xi_\varepsilon}.\label{Frformula}
\end{align}
This expression has the proper $2\pi$-periodicity in the superconducting phase, since $\eta\mapsto\eta+\pi$ and $\varepsilon\mapsto-\varepsilon$ if $\phi_1$ or $\phi_2$ is incremented by $2\pi$.

We evaluate $F(\bm{r})$ in a wide SNS junction, at a point $\bm{r}=(x_0,y_0)$ far from the lateral boundaries. A mode passing through this point at an angle $\theta$ relative to the $y$-axis has Aharonov-Bohm phase
\begin{align}
&\frac{2e}{\hbar}\int_{S_1}^{\bm{r}} \bm{A}\cdot d\bm{l}=\frac{(y_0+L/2)}{l_m^2}[2x_0-(y_0+L/2)\tan\theta],\nonumber\\
&\gamma=\frac{2e}{\hbar}\int_{S_1}^{S_2} \bm{A}\cdot d\bm{l}=\frac{2L}{l_m^2}(x_0-y_0\tan\theta),
\end{align}
so that the phase shift \eqref{etadef} is given by
\begin{equation}
\eta=\bar{\phi}-\frac{2x_0y_0}{l_m^2}+\frac{y_0^2+\tfrac{1}{4}L^2}{l_m^2}\tan\theta,\;\;\bar{\phi}=\tfrac{1}{2}(\phi_1+\phi_2).
\end{equation}

\begin{figure}[tb]
\centerline{\includegraphics[width=0.7\linewidth]{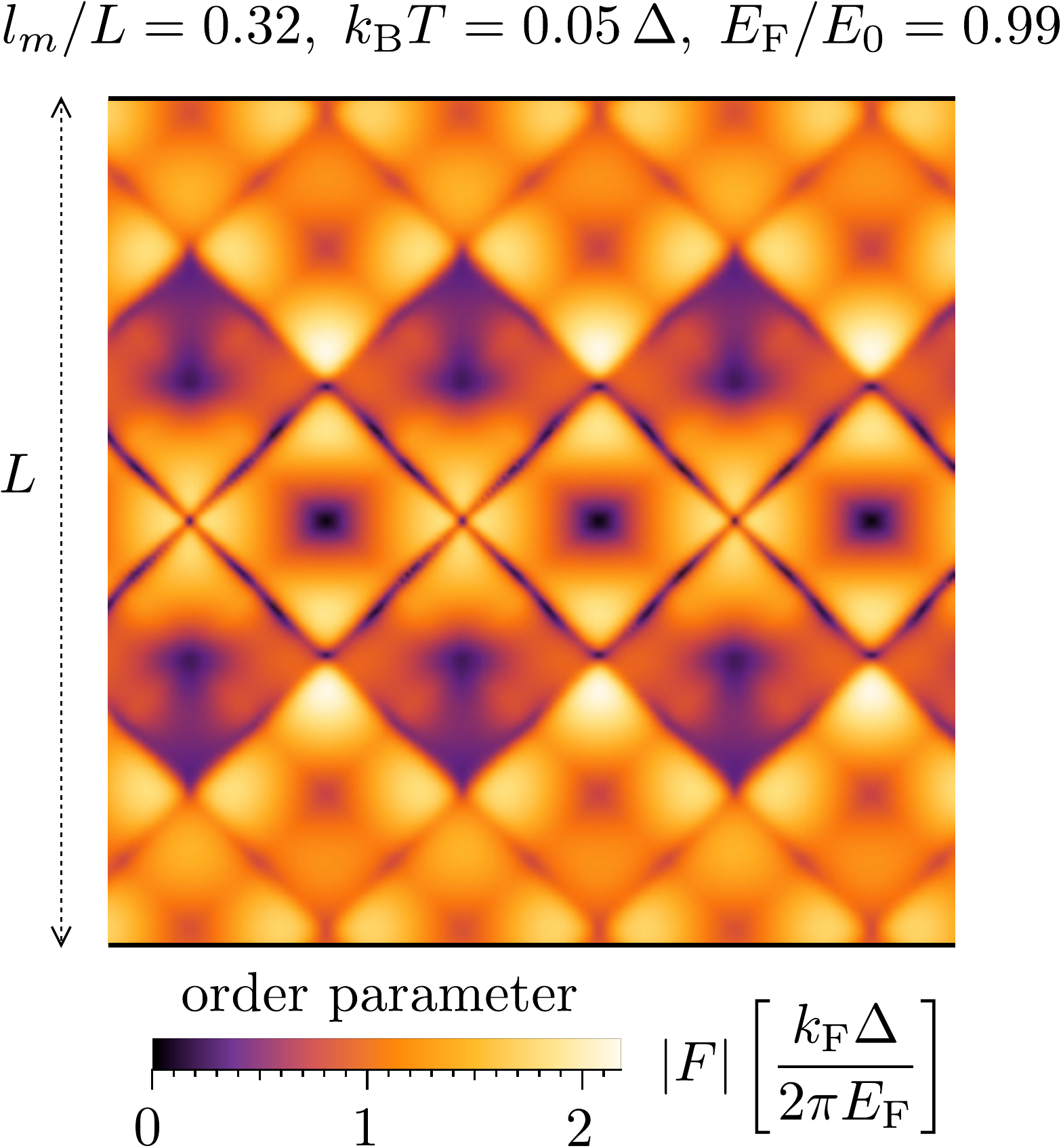}}
\caption{Absolute value of the superconducting order parameter $F(\bm{r})$, calculated from Eq.\ \eqref{Fresult}. Current vortices and antivortices in Fig.\ \ref{fig_vortex_lowT} correspond to local minima of $|F|$.
}
\label{fig_orderparameter}
\end{figure}

For the warped Fermi surface of a square lattice (unit lattice constant, see Sec.\ \ref{description}) we have
\begin{align}
&\tan\theta=\frac{\sin k_x}{\Xi},\;\;v_y=\frac{E_0\Xi}{2\hbar},\\
&\psi=\phi_1-\phi_2-\frac{2L}{l_m^2}\left(x_0-\frac{y_0}{\Xi}\sin k_x\right),\\
&\Xi=\sqrt{1-(\cos k_x+2E_{\rm F}/E_0-2)^2}.
\end{align}
The order parameter then results from the integral
\begin{align}
&F(\bm{r})=\frac{\Delta}{2\pi E_0}e^{-i\bar{\phi}}\exp(2ix_0y_0/l_m^2)\int_{-k_{\rm F}}^{k_{\rm F}} dk_x\, \frac{1}{\Xi}\left|\sin(\psi/2)\right|\nonumber\\
&\quad\times\tanh\left(\frac{\Delta\cos(\psi/2)}{2k_{\rm B}T}\right)\exp\left(-\frac{i(y_0^2+\tfrac{1}{4}L^2)}{ l_m^2 \Xi}\sin k_x\right),\label{Fresult}
\end{align}
with $k_{\rm F}={\rm arccos}\,(1-2E_{\rm F}/E_0)$. The resulting 2D lattice structure is shown in Fig.\ \ref{fig_orderparameter}, corresponding to the current vortex lattice of Fig.\ \ref{fig_vortex_lowT}.

\section{Two-dimensional lattice structure of the density of states}
\label{AppDOS}

To complete the picture, we also demonstrate the development of a 2D lattice structure in the density of states. The states at $\pm\varepsilon$ contribute $|\Psi(\bm{r})|^2[\delta(E+\varepsilon)+\delta(E-\varepsilon)]$ to the local density of states $\rho(\bm{r},E)$. The total contribution is
\begin{align}
\rho(\bm{r},E)&=\int \frac{dk_x}{2\pi}\bigl(|u(\bm{r})|^2+|v(\bm{r})|^2\bigr)\sum_{\sigma=\pm}\delta(E-\sigma\varepsilon)\nonumber\\
&=\int \frac{dk_x}{2\pi}\frac{\Delta}{\hbar v_{y}}|\sin(\psi/2)|\sum_{\sigma=\pm}\delta\bigl(E-\sigma\Delta\cos(\psi/2)\bigr).
\end{align}

\begin{figure}[tb]
\centerline{\includegraphics[width=0.7\linewidth]{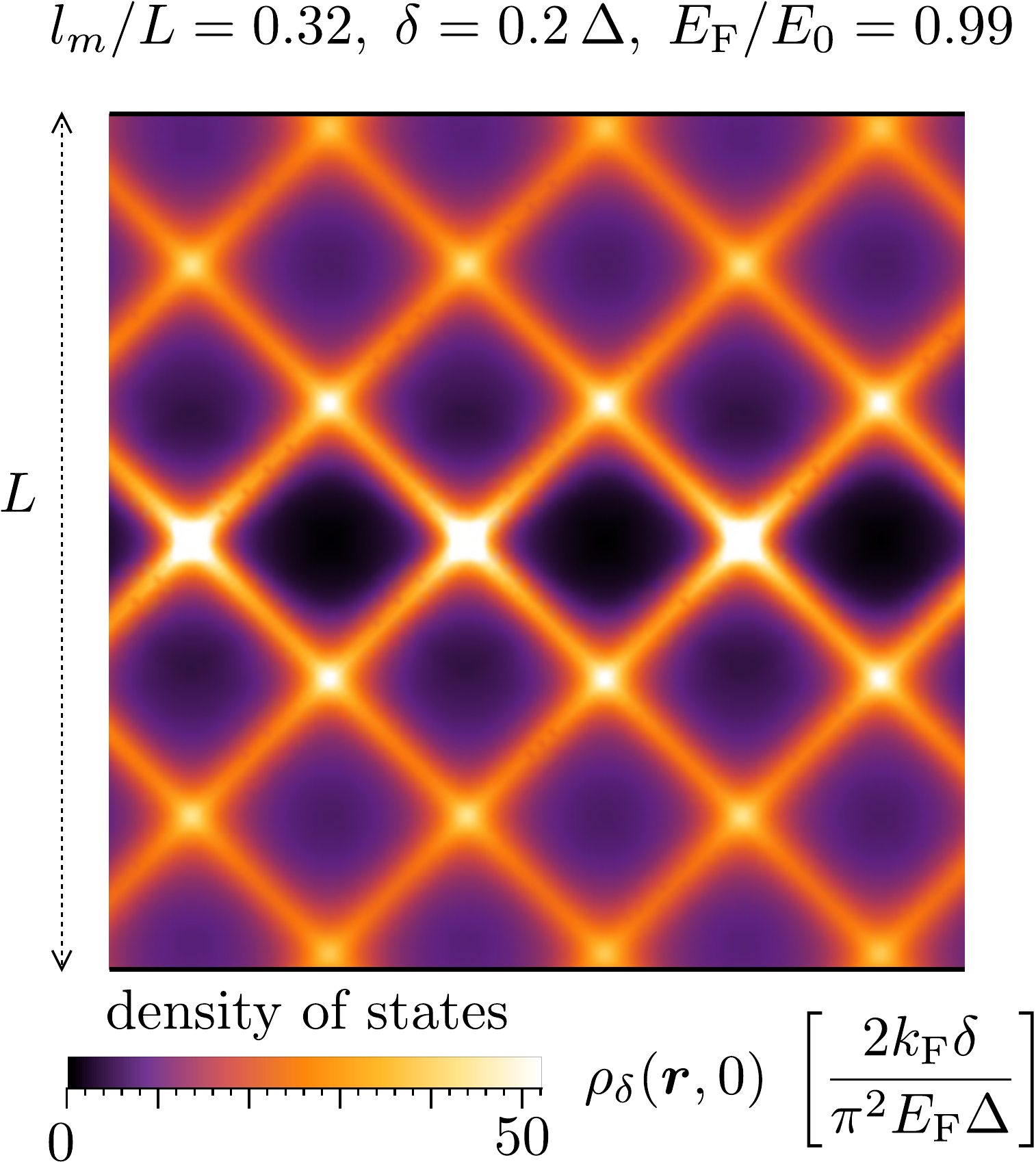}}
\caption{Local density of states $\rho_\delta$ at the Fermi level (with a Lorentzian broadening $\delta$), calculated from Eq.\ \eqref{DOSresult}. Current vortices and antivortices in Fig.\ \ref{fig_vortex_lowT} correspond to local maxima and minima of $\rho_\delta$.
}
\label{fig_DOS}
\end{figure}

We regularize the delta function by introducing a Lorentzian broadening $\delta$,
\begin{equation}
\rho_\delta(\bm{r},E)=\int \frac{dk_x}{2\pi}\frac{\Delta}{\hbar v_{y}}\sum_{\sigma=\pm}
\frac{(\delta/\pi)|\sin(\psi/2)|}{\delta^2+\bigl(E-\sigma\Delta\cos(\psi/2)\bigr)^2}.
\end{equation}
At the Fermi level, $E=0$, we evaluate
\begin{equation}
\rho_\delta(\bm{r},0)=\frac{2\delta}{\pi^2 E_0\Delta}\int_{-k_{\rm F}}^{k_{\rm F}} dk_x\,\frac{\Xi^{-1}|\sin(\psi/2)|}{(\delta/\Delta)^2+\cos^2(\psi/2)}.\label{DOSresult}
\end{equation}
The resulting 2D lattice is shown in Fig.\ \ref{fig_DOS}.


\begin{thebibliography}{99}
\bibitem{note1} At grazing incidence angles $|\theta|\rightarrow\pi/2$ the short-junction criterion is more stringent than $L\ll\xi$, we require $L\ll\xi\cos\theta$. In the analytics we ignore this complication, but it is fully incorporated in the numerics.
\bibitem{note2} The $1/B^2$ decay of the critical current for a circular Fermi surface follows upon numerical integration of Eq.\ \eqref{IcGamma} with $\rho(\theta)=1$, see Fig.\ \ref{fig_Fraunhofer}. Unlike the $1/\sqrt{B}$ decay for a square Fermi surface, we have not managed to derive the $1/B^2$ decay analytically.
\end{thebibliography}
\end{document}